\documentstyle[]{mn}
\input psfig.tex 

\def\be{\begin{equation}}
\def\ee{\end{equation}}

\newif\ifAMStwofonts
\ifoldfss

  \ifCUPmtlplainloaded \else
    \NewTextAlphabet{textbfit} {cmbxti10} {}
    \NewTextAlphabet{textbfss} {cmssbx10} {}
    \NewMathAlphabet{mathbfit} {cmbxti10} {} % for math mode
    \NewMathAlphabet{mathbfss} {cmssbx10} {} %  "   "    "
  \fi
  \ifAMStwofonts
    \ifCUPmtlplainloaded \else
      \NewSymbolFont{upmath} {eurm10}
      \NewSymbolFont{AMSa} {msam10}
      \NewMathSymbol{\upi}     {0}{upmath}{19}
      \NewMathSymbol{\umu}     {0}{upmath}{16}

      \NewMathSymbol{\upartial}{0}{upmath}{40}
      \NewMathSymbol{\leqslant}{3}{AMSa}{36}
      \NewMathSymbol{\geqslant}{3}{AMSa}{3E}

       \let\le=\leqslant
       \let\ge=\geqslant
    \fi
  \fi
\fi % End of OFSS

\ifnfssone
  \newmathalphabet{\mathit}
  \addtoversion{normal}{\mathit}{cmr}{m}{it}
  \addtoversion{bold}{\mathit}{cmr}{bx}{it}
  \newmathalphabet{\mathbfit} % math mode version of \textbfit{..}
  \addtoversion{normal}{\mathbfit}{cmr}{bx}{it}
  \addtoversion{bold}{\mathbfit}{cmr}{bx}{it}
  \newmathalphabet{\mathbfss} % math mode version of \textbfss{..}
  \addtoversion{normal}{\mathbfss}{cmss}{bx}{n}
  \addtoversion{bold}{\mathbfss}{cmss}{bx}{n}
  \ifAMStwofonts
    \ifCUPmtlplainloaded \else
      \UseAMStwoboldmath
      \makeatletter
      \new@mathgroup\upmath@group  
      \define@mathgroup\mv@normal\upmath@group{eur}{m}{n}
      \define@mathgroup\mv@bold\upmath@group{eur}{b}{n}
      \edef\UPM{\hexnumber\upmath@group}
      \new@mathgroup\amsa@group
      \define@mathgroup\mv@normal\amsa@group{msa}{m}{n}
  \edef\AMSa{\hexnumber\amsa@group}
      \makeatother
      \mathchardef\upi="0\UPM19
      \mathchardef\umu="0\UPM16
      \mathchardef\upartial="0\UPM40
      \mathchardef\leqslant="3\AMSa36
      \mathchardef\geqslant="3\AMSa3E

       \let\le=\leqslant
       \let\ge=\geqslant 
    \fi
  \fi
\fi % End of NFSS release 1

\ifnfsstwo
  \DeclareMathAlphabet{\mathbfit}{OT1}{cmr}{bx}{it}
  \SetMathAlphabet\mathbfit{bold}{OT1}{cmr}{bx}{it}
  \DeclareMathAlphabet{\mathbfss}{OT1}{cmss}{bx}{n}
  \SetMathAlphabet\mathbfss{bold}{OT1}{cmss}{bx}{n}
  \ifAMStwofonts
    \ifCUPmtlplainloaded \else
      \DeclareSymbolFont{UPM}{U}{eur}{m}{n}
      \SetSymbolFont{UPM}{bold}{U}{eur}{b}{n}
      \DeclareSymbolFont{AMSa}{U}{msa}{m}{n}
      \DeclareMathSymbol{\upi}{0}{UPM}{"19}
      \DeclareMathSymbol{\umu}{0}{UPM}{"16}
      \DeclareMathSymbol{\upartial}{0}{UPM}{"40}
      \DeclareMathSymbol{\leqslant}{3}{AMSa}{"36}
      \DeclareMathSymbol{\geqslant}{3}{AMSa}{"3E}

       \let\le=\leqslant
  \let\ge=\geqslant
    \fi
  \fi
\fi % End of NFSS release 2
   
\ifCUPmtlplainloaded \else
  \ifAMStwofonts \else % If no AMS fonts
    \def\upi{\pi}
    \def\umu{\mu}
    \def\upartial{\partial}
   \fi
\fi

\begin{document}

\title[Optimizing the yield of Sunyaev-Zel'dovich  cluster surveys] {Optimizing the yield of Sunyaev-Zel'dovich cluster surveys}

\author[Battye and Weller]
{Richard A. Battye$^{1}$\thanks{rbattye@jb.man.ac.uk}, Jochen Weller$^{2,3}$\thanks{jweller@fnal.gov}\\
$^{1}$Jodrell Bank Observatory, School of Physics \& Astronomy,
University of Manchester, Macclesfield, Cheshire SK11 9DL, UK. \\
$^{2}$NASA/Fermilab Astrophysics Group, Fermi National Accelerator Laboratory, Batavia IL 60510-0500, USA.\\
$^{3}$Department of Physics and Astronomy, University College London, London WC1E 6BT, UK.
}

\date{Received
**insert**; Accepted **insert**}

\pagerange{\pageref{firstpage}--\pageref{lastpage}} \pubyear{}

\maketitle
\label{firstpage}

\begin{abstract}

We consider the optimum depth of a cluster survey selected using the Sunyaev-Zel'dovich effect. By using simple models for the evolution of the cluster mass function and detailed modeling for a variety of observational techniques, we show that the optimum survey yield is achieved when the average size of the clusters selected is close to the size of the telescope beam. For a total power measurement, we compute the optimum noise threshold per beam as a function of the beam size and then discuss how our results can be used in more general situations. As a by-product we gain some insight into what is the most advantageous instrumental set-up. In the case of beam switching observations one is not severely limited if one manages to set the noise threshold close to the point which corresponds to the optimum yield. By defining a particular reference configuration, we show how our results can be applied to interferometer observations.  Considering a variety of alternative scenarios, we discuss how robust our conclusions are to modifications in the cluster model and cosmological parameters. The precise optimum is particularly sensitive to the amplitude of fluctuations  and the profile of the gas in the cluster.  
\end{abstract}


%%%%%%%%%%%%%%%%%%%%%%%%%%%%%%%%%%%%%%%%%%%%%%%%%%%%%%%%%%%%%%%%%%%%%%%%%%
%%%%%%%%%%%%%%%%%%%%%%%%%%%%%%%%%%%%%%%%%%%%%%%%%%%%%%%%%%%%%%%%%%%%%%%%%%

\section{Introduction}
\label{intro}

The Sunyaev-Zel'dovich (SZ) effect is due to the re-scattering of  cosmic microwave background (CMB) photons by the hot electrons situated in intracluster media~\cite{SZ}. It has been realized that it should provide a reliable mechanism for selecting a sample of clusters~\cite{HMH,HHM} and a number of instruments sensitive to the microwave frequency range (15GHz-350GHz) are under construction to perform this task~\cite{AMI,SZA,AMIBA,BOLOCAM,OCRA,SPT,ACT}. Since the cluster mass function is sensitive to cosmological parameters~\cite{TC,OB,SCO,B,VL,ECF} such surveys are likely to be an important part of the program to probe the late-time history of the universe, searching for the origin of the apparently accelerating expansion of the Universe. 

The number of clusters as a function of redshift and flux is a strong function of $\Omega_{\rm m}$, the mass density of the universe in units of the critical density, and $\sigma_8$, the normalization of the matter power spectrum on scales of $8h^{-1}{\rm Mpc}$, and is a weaker function of the Hubble constant $H_0=100h\,{\rm km}\,{\rm sec}^{-1}{\rm Mpc}^{-1}$, the equation of state $w_0$ of the dark energy, Q, and its evolution $w_1$. Here, $w_0$ and $w_1$ are defined by $P_{\rm Q}/\rho_{\rm Q}=w=w_0+w_1z$. It has been shown that the degeneracies within the space of cosmological parameters are such that cluster surveys are complementary to those which might be obtained using observations of Type Ia Supernovae at high redshift~\cite{WBK,BW}.

Since the distribution of the clusters obeys Poisson statistics the errorbars deduced on the cosmological parameters will be very sensitive to the number of clusters found in a given survey. Therefore, we anticipate that finding the largest number of clusters possible for a given integration time  will be close to giving the best constraints on cosmological parameters, although this may also weakly depend on their distribution with redshift. Clearly a survey which will find $\sim 1000$ clusters will constrain cosmological parameters much better than one which finds $\sim 100$. Having a much larger sample would also seem desirable in terms of reducing the systematic errors due to the poorly understood gas physics. Fortunately, we will show that, by  assuming we can model the gas physics well, one can get to within a factor of two of the optimal situation for a total power measurement. This is more tricky in more complicated experimental situations such as beam switching whereby an ill-judged selection of the survey depth can lead to one finding very few clusters at all.

The goal of this paper, therefore,  is to discuss issues related to optimizing the yield of a given survey; the optimum being defined as the survey in which the largest number of objects are found in a given amount of observing time. This is a question one is always faced with in any kind of astronomical measurement: how long does one integrate for on a given patch of sky? Obviously the shorter the required integration time for individual patches, the more sky area can be covered. However, if one doesn't integrate long enough only the very brightest objects will be found. Clearly, one is always guided in such situations by one's expectations of what is there to be observed.

For the present case of clusters, one can attempt to make a simple estimate by assuming (1) the clusters are point-like objects in the telescope beam, (2) their comoving number density is constant and (3) the Universe is Euclidean. In this case the number of objects with flux density, $S$, greater than the limiting flux density of a survey $S_{\rm lim}$ is $N(S>S_{\rm lim})\propto \Delta\Omega/S^{3/2}_{\rm lim}$, where $\Delta\Omega$ is the angular coverage. Since for fixed total integration time $S_{\rm lim}\propto \sqrt{\Delta\Omega}$, one can deduce that $N\propto \Delta\Omega^{1/4} \propto S_{\rm lim}^{1/2}$, and hence a survey with largest possible area (that is, the whole sky) would be optimal in this case. Under these assumptions the survey would be flux density limited and hence an instrument with the largest possible collecting area would be optimal in terms of making the largest number of detections in the shortest period of time.

Unfortunately, in most cases clusters are unlikely to be point-like, nor is their comoving density likely to be constant due to evolution and, of course, the Universe is not Euclidean. As we will describe in section~\ref{distn}, the distribution of clusters as a function of mass and redshift can be modelled in a sensible way for a given set of cosmological parameters. However, the key to understanding the issue at hand here is how the limiting mass of a survey is related to the instrumental noise and the angular resolution of the telescope.

\begin{figure} 
\setlength{\unitlength}{1cm}
\centerline{\psfig{file=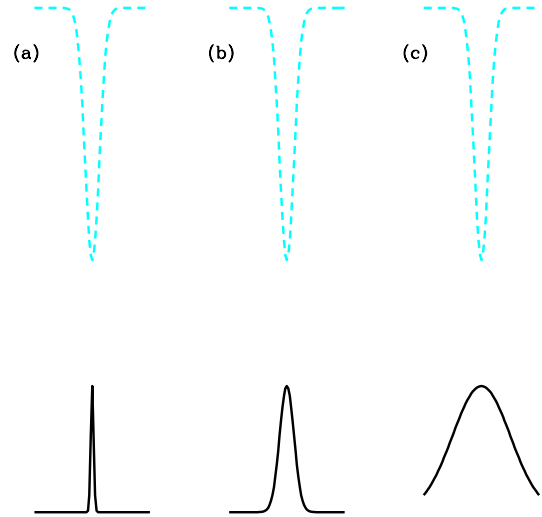,width=8cm,height=8cm}}
\caption{The top row is the negative cluster profile for an
  observation frequency of less than 217GHz. The bottom row are three
  hypothetical beam profiles: (a) the beamsize is much smaller than the
  cluster and the observation resolves it. In this case the
  observation is surface brightness temperature limited. (b) the
  beamsize exactly fits the cluster profile. This is the optimal
  situation. (c) the beamsize is larger than the cluster profile. In
  this case the observation is flux density limited.} 
\label{fig:beams}
\end{figure}

Shallow surveys, as suggested by the simple estimate above, will only find clusters which are very large and close by, and such clusters cover many beam areas. Since the same signal is spread over a number of beam areas, such objects can dip below the detection threshold  even though they are very large. It is clear that this effect will compete with that due to Euclidean source counts in our initial estimate, leading to an optimum at a particular value of the noise and angular coverage. We will show that this optimum will depend on the size of the beam: larger beam sizes, say $8^{\prime}$, will require a deeper\footnote{We define deep to mean a low noise level in terms of temperature.} survey to achieve the optimum, whereas a much shallower survey will be optimal for much smaller beam sizes, say $1^{\prime}$. Essentially the optimal situation is for most of the clusters in the survey to be the same size as the beam as explained in Fig.~\ref{fig:beams}.

Once the beam is sufficiently small to resolve all the clusters, the survey becomes surface brightness temperature limited, rather than being flux density limited, and increasing the collecting area no longer acts to  increase the rate of detection. This illustrates an interesting feature of cluster survey, which is in contrast to intuition one might have from considering standard point sources and, as we shall see, leads to some other somewhat counter-intuitive conclusions.

In section~\ref{mass} we will discuss how to derive a relation for the limiting mass of the survey as a function of redshift and the limiting flux, incorporating the effects of a Gaussian beam profile. Our initial motivation is to model idealized total power measurements using a single dish with a circular aperture. We will allow the observations to be smoothed using a window function which is exactly tuned to every possible cluster size above the maximum resolution of the telescope. Under these assumptions we show in section~\ref{total} that for a particular minimum beam size there exists a unique value of the noise which corresponds to the optimum in terms of the total of clusters detected, and for all beam sizes less than around $2^{\prime}$ (FWHM) this is equivalent in terms of the number of clusters found in the survey, although the precise optimum noise will be different for different beam sizes.

Accurate total power measurements are only feasible in space or at a very dry site (for example, the South Pole) on the earth's surface, mainly due to the effects of water vapour. In section~\ref{real} we discuss how our calculations can be adapted to situations closer to those experienced in reality. In particular, we will illustrate how our calculations can be applied to techniques which minimize the effects of the atmospheric contamination which is ubiquitous at the relevant frequencies at most sites. These techniques involve either use of a beam switch or an interferometric array. For the most part our conclusions are similar to those for a total power experiment, although more care is required in interpreting them.

Obviously our goal has some clear dangers since one is, by necessity, required to assume a specific distribution of clusters. In situations where the source population is completely unknown, attempts at optimizing observations in this way are completely useless! However, the distribution and precise characteristics of clusters are becoming much better understood; something which is only likely to improve as observations progress. We have, therefore, used the so-called concordance cosmological parameters~\cite{OS} and have attempted to model most of the relevant features of the distribution of clusters. In section~\ref{robust} we test the robustness of our predictions to variations in the cosmological parameters and in the cluster model. Small variations do not affect the optimum significantly but, not surprisingly, large variations substantially change our predictions. We quantify this in terms of the parameters. Of particular concern is the uncertainty in the detailed modelling of the poorly understood  gas physics within the cluster. It is clear that more sophisticated modelling of this physics would be required before we can have full confidence in our predicted optima; they should, nonetheless, provide a useful guide.

For the purposes of this paper, we will ignore the effects of confusion due to radio and infra-red (IR) point sources, the CMB anisotropies and that due the distribution of clusters themselves which are below the flux limit. These will be discussed in a future publication. Suffice to say that the limiting flux densities  and temperatures that we suggest here are compatible with confusion being negligible except for the most extreme cases. Very much deeper surveys run the risk of becoming confusion limited. We note that some of our conclusions concerning the most advantageous instrumental setup do not take into account possible strategies to mitigate problems which might occur due to confusion backgrounds.

We should note that our optimization  are only of relevance to ground-based experiments specifically designed to perform high resolution CMB measurements and SZ surveys at a single frequency. For general purpose, multi-frequency instruments such as PLANCK, the observational strategy (in the case of PLANCK, an all sky survey) has been designed to take into account a wide range of possible applications, not just the ability to detect the SZ in clusters.

Throughout this paper we will work in natural units where $c=\hbar=k_{\rm B}=1$. Standard conversion factors can be found, for example, in Kolb \& Turner (1989).

\section{Modelling the cluster distribution}
\label{distn}

Except in section~\ref{robust} where we test the robustness of our predictions, we will use a fiducial cosmological model which is a flat $\Lambda$CDM model with $(\Omega_{\rm m},\sigma_8,h,w_0,w_1)=(0.3,0.9,0.72,-1,0)$. This model is compatible with recent CMB~\cite{WMAP}, Large-Scale Structure (LSS)~\cite{2dfa,2dfb},  Hubble constant~\cite{HUBBLE} and weak lensing measurements~\cite{CHL}. The precise value of $\sigma_8$ is currently under close scrutiny and is very important in making predictions of cluster observations since the number of objects found is extremely sensitive to its value. We should note that our chosen value is larger than some other values quoted in the literature~\cite{ASF} and we will return to this issue in section~\ref{robust}. Where necessary we use a baryon density which satisfies current estimates~\cite{WMAP} and a spectral index of initial density fluctuations $n_{\rm S}=1$. Only very extreme changes from these two parameters can lead to significantly different results.

We will use the mass function deduced from the suite of N-body simulations~\cite{EVRARD} performed by the VIRGO collaboration. The comoving number density of objects at redshift $z$ in the mass range $M$ to $M+dM$  is given by 
\be 
{dn\over dM}(z,M)=-0.22{\rho_{\rm m}(t_0)\over M\sigma_M}{d\sigma_M\over dM}\exp\left\{-|A(z,M)|^{3.86}\right\}\,,
\ee
where $A(z,M)=0.73-\log[D(z)\sigma_M]$, $\rho_{\rm m}(t_0)$ is the matter density at the present day, $\sigma_M$ is the over density in a virialized region containing mass $M$ and  $D(z)$ is the normalized growth factor which is dependent on cosmological parameters. Here, $M$ is defined to be $M_{200}$, that inside a region with an overdensity of 200, see  Battye \& Weller (2003) for more details on how to apply this to cluster calculations.

The clusters are assumed to be spherical and in hydrodynamical equilibrium, in which case the virial mass, $M_{\rm vir}$, and virial radius, $R_{\rm vir}$, are related by 
\be
M_{\rm vir}={4\pi\over 3}\rho_{\rm crit}(z)\Delta_{\rm c}(z)R_{\rm vir}^3\,,
\ee
where $\rho_{\rm crit}(z)$ is the critical density at redshift $z$ and $\Delta_{\rm c}(z)$ is the relevant overdensity which can be computed from the spherical collapse model~\cite{PP,GG,L}. This  can be adapted to apply to models where $w\ne -1$~\cite{WS,BW}.

We will assume that the electron density weighted temperature $\langle T_{\rm e}\rangle_n$ can be related to $M_{\rm vir}$ under the assumption that the cluster is isothermal via
\begin{eqnarray} 
{M_{\rm vir}\over 10^{15}h^{-1}M_{\odot}}&=&\left({\langle T_{\rm e}\rangle_n\over T_\star\,{\rm keV}}\right)^{3/2}\left[\Delta_{\rm c}(z)E(z)^2\right]^{-1/2}\cr
& &\times\left[1+(1+3w){\Omega_{Q}(z)\over \Delta_{\rm c}(z)}\right]^{-3/2}\,,
\label{mt}
\end{eqnarray}
where the Hubble parameter $H(z)=H_0E(z)$ and $\Omega_{Q}(z)$ is the density of the dark energy component relative to the critical density at redshift $z$. We will use $T_\star\approx 1.6$ as the normalization factor for most of our discussion, but will consider the effects of other values of $T_\star$ in section~\ref{robust}.

The profile of the hot gas $\zeta(\theta)$, whose mass is related to the total mass of the cluster by $M_{\rm gas}=f_{\rm gas}M_{\rm vir}$, is modelled using the isothermal $\beta$-model~\cite{BETA} with $\beta=2/3$ as suggested by observations,
\be 
\zeta(\theta)=\left(1+{\theta^2\over \theta_{\rm c}^2}\right)^{-1/2}{\tan^{-1}\left[\left({\alpha^2-\theta^2/\theta_{\rm c}^2\over 1+\theta^2/\theta_{\rm c}^2}\right)^{1/2}\right]\over\tan^{-1}\alpha}\,,
\ee
where $\theta_{\rm c}$ is the apparent angular size of the cluster core, $\alpha=\theta_{\rm vir}/\theta_{\rm c}$ and $\theta_{\rm vir}$ is related to $R_{\rm vir}$ via the angular diameter distance $d_A(z)$. We will assume that the gas fraction $f_{\rm gas}$ is universal and redshift independent with value of 0.12~\cite{GAS}. Essentially, uncertainties in the global value of $f_{\rm gas}$ can be absorbed into the mass-temperature  (\ref{mt}), via modifications in the $T_\star$. Variations in $f_{\rm gas}$ from cluster to cluster are more difficult to deal with.

The total number of clusters with $z<z_{\rm max}$ is computed using 
\be
N_{\rm tot}(z_{\rm max})=\Delta\Omega\int_0^{z_{\rm max}}dz\,{dV\over dzd\Omega}\int_{M_{\rm lim}(z)}^{\infty}dM\,{dn\over dM}\,,
\label{number}
\ee
where $\Delta\Omega$ is the angular coverage and $dV/(dzd\Omega)$ is the FRW comoving volume element. $M_{\rm lim}(z)$ is the limiting mass of the survey which one can relate to the limiting flux density $S_{\rm lim}$ as discussed in the section~\ref{mass}. We will use $z_{\rm max}=1.5$ since we are interested in clusters whose redshifts can be found using standard techniques. At present it is difficult to accurately deduce cluster redshifts which are $\sim 1$. We optimistically assume that,  by the time these SZ surveys have been performed, techniques will have been developed to at least deduce photometric redshifts routinely out to $z=1.5$. There are expected to be very few clusters with $z>1.5$. In principle our calculations could easily be modified to optimize the  number of clusters between any two redshifts as desired.

\section{Computing the limiting mass}
\label{mass}

\begin{figure} 
\setlength{\unitlength}{1cm}
\centerline{\psfig{file=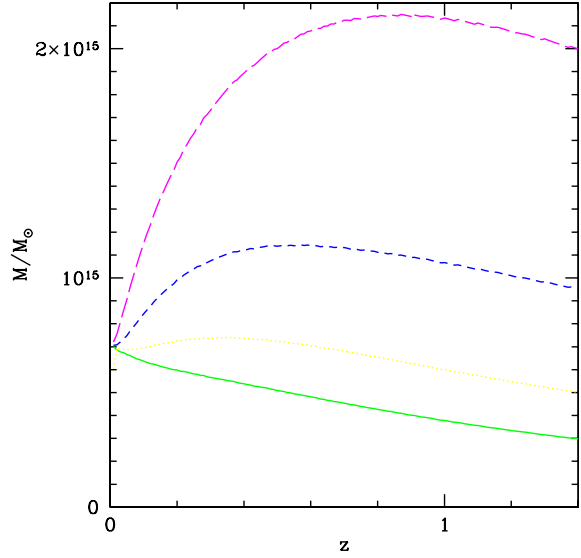,width=8cm,height=8cm}}
\caption{The limiting mass for a family of surveys with $f=30{\rm GHz}$ and $S_{\rm lim}/\theta_{\rm FWHM}^2=500\mu{\rm Jy}\,{\rm arcmin}^{-2}$, that is, a family with the same system temperature, but different beam sizes. Each of the curves uses a beam of fixed resolution placed over the centre of the cluster. The solid line has $\theta_{\rm FWHM}=1^{\prime}$, the dotted line $\theta_{\rm FWHM}=2^{\prime}$, the short-dashed line $\theta_{\rm FWHM}=4^{\prime}$ and the long dashed line $\theta_{\rm FWHM}=8^{\prime}$. The shape of the $8^{\prime}$ beam approximately follows that of $d^{6/5}_{\rm A}$ which is indicative of the point source approximation and with $\theta_{\rm FWHM}=1^{\prime}$ it is more  difficult to detect clusters which are close by (Bartlett 2000), when compared to those at higher redshifts. Using the fiducial model $N_{\rm tot}(1.5)=0.33, 8.3\times 10^{-2}, 1.6\times 10^{-2}, 2.4\times 10^{-3}$ per ${\rm deg^2}$ for $\theta_{\rm FWHM}=1^{\prime}$, $2^{\prime}$, $4^{\prime}$ and $8^{\prime}$ respectively. Note that each of the curves tends to the same non-zero value as described in Appendix A.}
\label{fig:mlim1}
\end{figure}

The flux density per beam due to the Sunyaev-Zel'dovich effect is given by
\be
S=2f^2T_{\rm CMB}h(x)Y\,,
\ee 
where $T_{\rm CMB}$ is the CMB temperature, $Y$ is the Compton  y-distortion integrated over the beam profile and $h(x)$ is given by 
\be
h(x)={x^2e^{x}\over (e^{x}-1)^2}\left[{x\over \tanh(x/2)}-4\right]=-2+{\cal O}(x)\,,
\ee
where $x=2\pi f/T_{\rm CMB}\approx f/56.8{\rm GHz}$, $f$ is the frequency of observation and $\lambda$ is the corresponding wavelength.
The limiting virial mass of the survey can be related to the flux density per beam~\cite{BW} by 
\be 
\label{smt}
S={2f^2 T_{\rm CMB}\langle T_{\rm e}\rangle_n\sigma_{\rm T}h(x)\over m_{\rm e} d_{A}^2}{f_{\rm gas}M_{\rm vir}\over \mu_{\rm e}m_{\rm p}}{\cal I}(B, \zeta)\,,
\ee
where 
\be 
{\cal I}(B,\zeta)={\int d\Omega B(\theta)\zeta(\theta)\over \int  d\Omega\zeta(\theta)}\,,
\label{i}
\ee
$\sigma_{\rm T}$ is the Thomson scattering cross-section, $\mu_{\rm e}\approx 1.143 $ is the mean molecular weight of the electrons and $m_{\rm e}, m_{\rm p}$ are the masses of the electron and the proton, respectively. The corresponding beam area is given by 
\be 
\Omega_{\rm b}=\int d\Omega B(\theta)\,.
\ee

The corresponding thermodynamic temperature can be computed using the standard formula
\be 
T^{\rm THERM}={g(x)S\over 2f^2h(x)\Omega_{\rm b}}\,,\label{st}
\ee
and
\be
g(x)={x\over\tanh(x/2)}-4\,,
\ee
is the frequency dependence of the SZ brightness temperature. We note that this quantity is not the peak of the SZ effect due to the cluster, but the average integrated over a region whose size is set by the instrumental beam size. For the purposes of our discussion we will define a non-standard temperature unit which includes this frequency dependence, allowing us, where desirable, to quote frequency independent limits in the subsequent discussion,
\be 
T={T^{\rm THERM}\over g(x)}={T_{\rm CMB}\langle T_{\rm e}\rangle_{n}\sigma_{\rm T}f_{\rm gas}M_{\rm vir}\over \mu_{\rm e}m_{\rm e}m_{\rm p}d_{\rm A}^2}{{\cal I}(B,\zeta)\over \Omega_{\rm b}}\,.
\ee
This can be related to the integrated $y$-distortion by $Y=T/T_{\rm CMB}$.
 
The function ${\cal I}(B,\zeta)\le \Omega_{\rm b}$ takes into account the effects of the beam $B(\theta)$  centred directly over the centre of the cluster. We will take this to be a Gaussian with full-width half-max (FWHM) $\theta_{\rm FWHM}$,
\be
B(\theta)=\exp\left(-{\theta^2\over 2\sigma_{\rm b}^2}\right)\,,
\ee
where $\sigma_{\rm b}=\theta_{\rm FWHM}/\sqrt{8\log 2}$ and $\Omega_{\rm b}=2\pi\sigma_{\rm b}^2$. It can be shown analytically (see Appendix A) that, under sensible assumptions for $\zeta(\theta)$, which apply to the isothermal $\beta$-model and other possible cluster profile functions,  the divergence in (\ref{smt}) at $z=0$  due to $d_A(z)$, is regularized by ${\cal I}$. Hence, $M_{\rm lim}(z=0)$ is non-zero; this is critical in the context of our subsequent discussion since it is this which mitigates against our earlier comments in the introduction on a shallow survey being optimal.

\begin{figure} 
\setlength{\unitlength}{1cm}
\centerline{\psfig{file=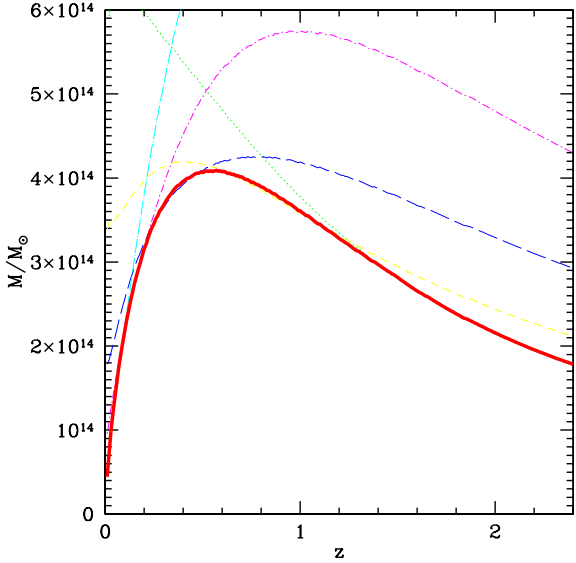,width=8cm,height=8cm}}
\caption{The solid line represents the limiting mass for $S_{\rm lim}^{\rm min}=500\mu{\rm Jy}$, $f=30{\rm GHz}$ and $\theta_{\rm FWHM}^{\rm min}=1^{\prime}$ computed using the algorithm discussed in the text allowing $N_{\rm b}$ to take any value. Included also are curves of fixed resolution with $\theta_{\rm FWHM}=1^{\prime}$ (dotted line), $2^{\prime}$ (short dashed line), $4^{\prime}$ (long dashed line), $8^{\prime}$ (short dashed-dotted line), $16^{\prime}$ (long dashed-dotted line). The limiting flux densities  have been adjusted so that $S_{\rm lim}=500\mu{\rm Jy}$, $1{\rm mJy}$, $2{\rm mJy}$, $4{\rm mJy}$ and $8{\rm mJy}$ respectively to model the effects of beam smoothing. Note that the solid curve represents the envelope of the others.}
\label{fig:mlim2}
\end{figure}

Our aim is to use relation (\ref{smt}) to deduce the limiting mass as a function of redshift for a survey with noise quantified by limiting flux density $S_{\rm lim}$ per beam. In order to do this we first consider the simple case where we do not take into account of any kind of beam smoothing and assume the cluster to be `detected' if the flux density in a beam which lies directly over its centre is greater than $S_{\rm lim}$. The beam correction ${\cal I}$ is a complicated function of $M_{\rm vir}$ and $z$, but (\ref{smt}) can be inverted to give $M_{\rm lim}$ as function of $z$ using the Newton-Raphson method. We have computed this for $f=30{\rm GHz}$ and the results are illustrated in Fig.~\ref{fig:mlim1} for a set of surveys which have constant $S_{\rm lim}/\theta_{\rm FWHM}^2$, that is, ones with the same brightness temperature sensitivity. We see that for large beams, $\theta_{\rm FWHM}\sim 8^{\prime}$, the shape of the curve reflects the $d_{A}^{6/5}$ dependence for small $z$ of the limiting mass expected in the point source approximation~\cite{BART}. In the absence of beam smoothing, much smaller beams give an altogether different shape since they find it more difficult to detect clusters which are very large, that is, those which are very close by. This led Bartlett (2000) to suggest that surveys performed  with very high resolution instruments should be very deep, contrary to the argument based on Euclidean source counts we presented earlier. However, one must also attempt to take into account the effects of beam smoothing and be careful to make comparisons with configurations of the same temperature sensitivity, as we have done here, rather than flux density sensitivity which is related to the overall collecting area. As we will see this leads to a conclusion somewhere in between these two extremes, strongly dependent on the resolution of the telescope.

In a diffraction limited observation,  the highest resolution beam size possible is $\theta_{\rm FWHM}^{\rm min}\propto\lambda/D$ where $D$ is the diameter of the telescope. Coarser resolution is possible by either physically under-illuminating the dish using appropriate feed horns, or by beam smoothing the resultant map. Let us consider the latter possibility of an increased beam size  $\theta_{\rm FWHM}=\sqrt{N_{\rm b}}\theta_{\rm FWHM}^{\rm min}$ whose area covers $N_{\rm b}$ copies of the minimum beam size, that is, $\Omega_{\rm b}=N_{\rm b}\Omega_{\rm b}^{\rm min}$. By adding the noise together for the $N_{\rm b}$ beams, we see that $T_{\rm lim}=T_{\rm lim}^{\rm min}/\sqrt{N_{\rm b}}$ where $T_{\rm lim}^{\rm min}$ is the noise temperature corresponding to the minimum beam size, and  the  noise level in terms of flux density increases to respect (\ref{st}). Hence, the limiting flux density becomes $S_{\rm lim}=\sqrt{N_{\rm b}}S_{\rm lim}^{\rm min}$ where $S_{\rm lim}^{\rm min}$ is the flux density limit equivalent to $T_{\rm lim}^{\rm min}$. This would correspond to taking a map with resolution determined by $\Theta_{\rm FWHM}^{\rm min}$ and convolving it with a Gaussian window function with FWHM $\sqrt{N_{\rm b}}\theta_{\rm FWHM}^{\rm min}$.

A simple, and unrealistic, assumption is that one can smooth the resulting map to the size of any cluster larger than the minimum beam size which might contribute to it, that is, $N_{\rm b}$ can take any value $\ge1$ adjusted to the size of the cluster in question. This represents the most one could hope to achieve from a particular observation. The number of beam areas covered by a cluster  weighted by the gas profile function is given by 
\be 
N_{\rm b}={\int \theta d\theta\,\zeta(\theta)\over \int \theta d\theta\,B(\theta)}\,.
\ee
This can be used in conjunction with the Newton Raphson method to compute the minimum limiting mass. This is shown in Fig.~\ref{fig:mlim2} as well as curves of fixed resolution  with  $N_{\rm b}=1,4,16,256$, which correspond to $\theta_{\rm FWHM}=(2^{\prime})^k$ and $N_{\rm b}=4^{k}$ with $k=0,1,2,3,4$. We have used  $S^{\rm min}_{\rm lim}=500\mu{\rm Jy}$, $f=30{\rm GHz}$ and $\theta_{\rm FWHM}^{\rm min}=1^{\prime}$. The minimum limiting mass clearly represents the envelope of the family of curves with $N_{\rm b}$ taking any value between 1 and $\infty$. This will be the basis for our idealized total power measurements discussed in section~\ref{total}

\begin{figure} 
\setlength{\unitlength}{1cm}
\centerline{\psfig{file=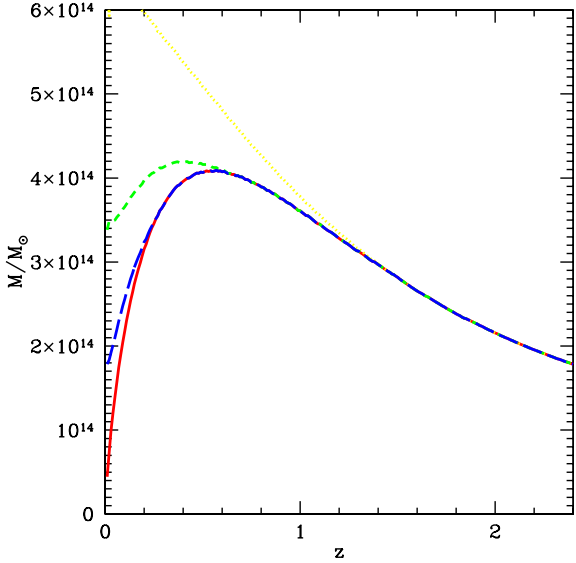,width=8cm,height=8cm}}
\caption{The solid line is the optimum limiting mass allowing for beam smoothing with any value of $N_{\rm b}$ for $S_{\rm lim}^{\rm min}=500\mu{\rm Jy}$, $f=30{\rm GHz}$, $\theta_{\rm FWHM}^{\rm max}=\infty$ and $\theta_{\rm FWHM}^{\rm min}=1^{\prime}$ and is the same as the solid line in Fig.~\ref{fig:mlim2}. The other curves represent the effects of imposing a finite value of $\theta_{\rm FWHM}^{\rm max}$ still allowing for any value of $N_{\rm b}$ up to this limit. The dotted line has $\theta_{\rm FWHM}^{\rm max}=1^{\prime}$ which is equivalent to the $1^{\prime}$ curve in Fig.~\ref{fig:mlim2},  the short-dashed line it is $2^{\prime}$ and for the long-dashed line it is $4^{\prime}$. It is clear that extending the beam much above $4^{\prime}$ will have little effect on the number of clusters found in the survey. Using the fiducial model $N_{\rm tot}(1.5)=0.33, 0.57, 0.67, 0.70$ per ${\rm deg}^2$ for $\theta_{\rm FWHM}^{\rm max}=1^{\prime}$, $2^{\prime}$, $4^{\prime}$ and $\infty$ respectively.}
\label{fig:mlim3}
\end{figure}

One might be prevented from using this beam smoothing procedure to detect objects very much larger than the minimum resolution, that is, there exists a maximum beam size $\theta_{\rm FWHM}^{\rm max}$ above which there should be no beam smoothing. This might be due, for example, to instrumental baselines. Moreover, in practice the value of $N_{\rm b}$ will  be required to take a discrete set of values (we note that this may not necessarily correspond to $N_{\rm b}$ taking integer values). One simple procedure that one might use is to take 4 beam areas from the highest resolution map to form a map with half the resolution; one which can then be iterated further to give an even lower resolution map. In this case $N_{\rm b}=4^{k}$ for some integer $k$.

\begin{figure} 
\setlength{\unitlength}{1cm}
\centerline{\psfig{file=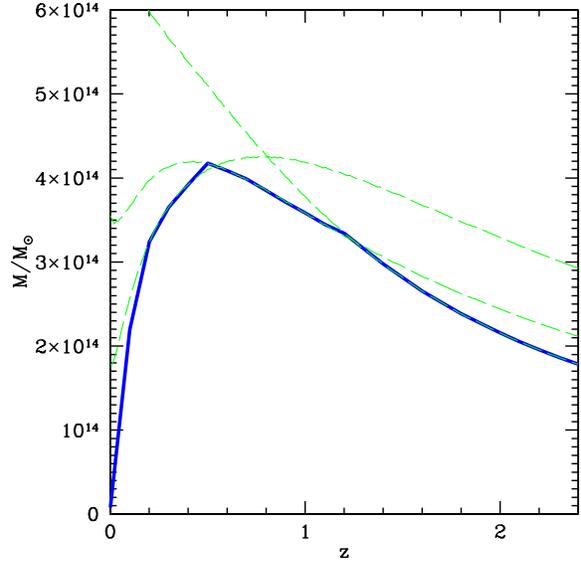,width=8cm,height=8cm}}
\caption{The solid line is the exact envelope of the dashed lines which are the same as the $\theta_{\rm FWHM}=1^{\prime}$, $2^{\prime}$ and $4^{\prime}$ from Fig.~\ref{fig:mlim2} which was computed by allowing $N_{\rm b}=4^{k}$ as described in the text. Notice that it is similar to, but not the same as, the solid line in Fig.~\ref{fig:mlim2} which allows any value of $N_{\rm b}$. It is unlikely that accurate inclusion of these effects of realistic beam smoothing will significantly alter the conclusions of this paper.}
\label{fig:mlim4}
\end{figure}

We investigate the effects of these two more realistic scenarios on the limiting mass in Figs.~\ref{fig:mlim3} and ~\ref{fig:mlim4}. In Fig.~\ref{fig:mlim3} we reduce $\theta_{\rm FWHM}^{\rm max}$ while still allowing any real value for $N_{\rm b}$, whereas in Fig.~\ref{fig:mlim4} we only allow $N_{\rm b}=4^{k}$, while keeping $\theta_{\rm FWHM}^{\rm max}=\infty$. We see that if $\theta_{\rm FWHM}^{\rm max}>4^{\prime}$, the limiting mass and, hence, the number of objects found in the survey, is not significantly affected by a finite value of $\theta_{\rm FWHM}^{\rm max}$. Moreover, it is clear that the discreteness effects introduced by considering only values of $N_{\rm b}$ which are multiples of 4 are unlikely to have a particularly strong effect on the number of clusters found in the survey since the solid curve in Fig.~\ref{fig:mlim4} is very close to that in Fig.~\ref{fig:mlim2}.

\section{Idealized total power measurements}
\label{total}

\begin{figure} 
\setlength{\unitlength}{1cm}
%\vskip -2cm
\centerline{\psfig{file=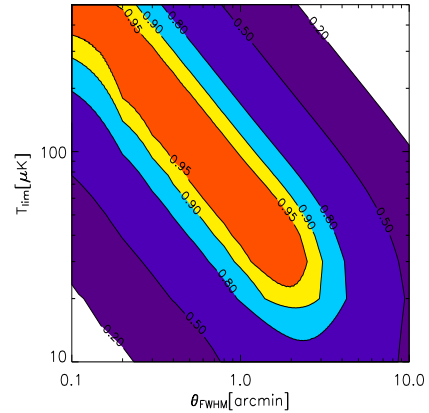,width=6.0cm,height=6.0cm}}
%\vskip -1cm
\caption{Contours of the predicted number of clusters computed using (\ref{ntot}) found per unit $C$ as a function of $\theta_{\rm FWHM}$ and $T_{\rm lim}$ per beam in the case of optimal beam smoothing. We have used $z_{\rm max}=1.5$. The contours represent 95\%, 90\%, 80\%, 50\% and 20\% of the maximum number that could be found with particular fixed values of $T_{\rm sys}$, $\eta$, $n_{\rm b}$, $t$ and $\Delta f$. Notice that, for $\theta_{\rm FWHM}<2^{\prime}$ there is a degenerate line $T_{\rm lim}=4T_{\sigma}\approx 55\mu{\rm K}/(\theta_{\rm FWHM}/1^{\prime})$ along which the survey would have optimal yield.}
\label{fig:tot1a}
\end{figure}

\begin{figure} 
\setlength{\unitlength}{1cm}
%\vskip 5cm
\centerline{\psfig{file=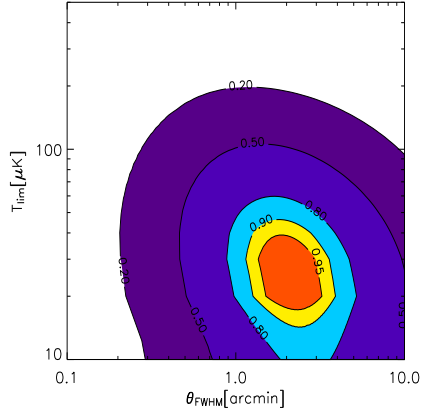,width=6.0cm,height=6.0cm}}
%\vskip -1cm
\caption{Equivalent to Fig.~\ref{fig:tot1a} without allowing for beam smoothing. There is a localized maximum near $\theta_{\rm FWHM}\approx 2^{\prime}$ and $T_{\rm lim}\approx 30\mu {\rm K}$. No degenerate line exists. We see that survey setups with $\theta_{\rm FWHM}<1^{\prime}$ find it difficult to find anything close to the optimal number of clusters, since they are unable to detect objects at low redshift without beam smoothing.}
\label{fig:tot1b}
\end{figure}

We will consider, initially, a simple single telescope observation characterized by the system temperature, $T_{\rm sys}$, the receiver bandwidth, $\Delta f$, the number of simultaneous beams, $n_{\rm b}$, generated by a focal plane array, and the diameter of the telescope $D$. The expected $1-\sigma$ instantaneous sensitivity is $T_{\rm inst} \approx T_{\rm sys}/(\eta\sqrt{\Delta f})$ where $\eta$ is the aperture efficiency of the telescope. In the rest of our discussion we shall use $\theta_{\rm FWHM}$ to refer to the minimum beam size $\theta_{\rm FWHM}^{\rm min}\approx\lambda/D$.

The question which we would like to answer is how this sensitivity can be deployed most effectively. The extremes are to point the telescope at the the same piece of sky for the whole period of integration, or to cover the whole sky with a very shallow survey. Of course, we are likely to do neither, but the noise temperature per beam of any survey, $T_{\sigma}$, and the area covered, $\Delta\Omega$, will satisfy 
\begin{eqnarray} 
{T_{\sigma}\over \sqrt{\Delta\Omega}}&=&\sqrt{4\log 2\over \pi}{T_{\rm sys}\over\theta_{\rm FWHM}\eta\sqrt{n_{\rm b}t\Delta f}}\cr &\approx& 0.94 {T_{\rm sys}\over \theta_{\rm FWHM}\eta\sqrt{n_{\rm b}t\Delta f}}\,,
\label{noise}
\end{eqnarray}
where $t$ is the integration time and we have assumed a Gaussian beam profile. Note that the temperatures, including the system temperatures, quoted here, unless explicitly stated otherwise,  are expressed in terms of the sensitivity to the SZ effect (see the earlier discussion of these non-standard temperature units).

This noise can be converted into the equivalent flux density limit via the standard formula 
\be
{S_{\sigma}\over\mu{\rm Jy}}=3.4\times 10^4h(x)\left({f\over {\rm GHz}}\right)^2
\left({T_{\sigma}\over\mu{\rm K}}\right)\theta_{\rm FWHM}^2\,\,.
\label{st1}
\ee
We will choose the detection threshold to be 4 times the noise, that is, $T_{\rm lim}=4T_{\sigma}$ and $S_{\rm lim}=4S_{\sigma}$. 

One can compute the number of objects for the family of surveys which satisfy (\ref{noise}) by substituting for $\Delta\Omega$ from (\ref{noise}) into (\ref{number}) to give 
\be
\label{ntot}
{N_{\rm tot}(z_{\rm max})\over CT_{\rm lim}^2\theta_{\rm FWHM}^2}=\int^{z_{\rm max}}_{0}dz\,{dV\over dzd\Omega}\int^{\infty}_{M_{\rm lim}(z)}dM{dn\over dM}\,,
\ee
where 
\begin{eqnarray}
C={\pi\over 64\log 2}{n_{\rm b}t \eta^2\Delta f\over T_{\rm sys}^2}\approx 0.07 {n_{\rm b}t\Delta f\over T_{\rm sys}^2}\,.
\end{eqnarray}
Here, $M_{\rm lim}(z)$ is now a function of $T_{\rm lim}$ and $\theta_{\rm FWHM}$. This expression shows the obvious dependencies on the parameters which we are taking to be constant $T_{\rm sys}$, $n_{\rm b}$, $t$, $\eta$ and $\Delta f$ : reducing $T_{\rm sys}$ and increasing $\Delta f$, $t$, $n_{\rm b}$, $\eta$  will lead to an increase in the number of clusters found.

The quantity 
\be 
{\cal M}_{\rm SD}={n_{\rm b}\eta^2\Delta f\over T_{\rm sys}^2}={n_{\rm b}\eta^2\Delta f\over (T_{\rm sys}^{\rm THERM})^2}[g(x)]^2\,,
\ee
can be considered as a figure of merit for a single dish instrument if $\theta_{\rm FWHM}<2^{\prime}$. Ignoring possible systematic effects it quantifies the number of objects an instrument will find for a fixed integration time, or conversely the time taken to find a fixed number of objects is $\propto{\cal M}_{\rm SD}^{-1}$. We see that once this limit on $\theta_{\rm FWHM}$ is saturated, the rate of making detection is independent of the overall collecting area, in contrast to the case for point sources. We also see that the survey goes from being flux density limited to surface brightness temperature limited at this saturation point. Given that currently available receivers have values of $T_{\rm sys}$ and $\Delta f$ which are close physical limits, one sees that the only way to improve the detection rate is by increasing $n_{\rm b}$. We note that the largest value of $n_{\rm b}$ is limited by the size of the focal plane of the telescope. 

An important consequence of an SZ survey being surface brightness temperature limited is that, for fixed values of $T_{\rm sys}$ and $\Delta f$, the optimal frequency to observe at would be the lowest possible, that is, $x\approx 0$, where the SZ effect is strongest in terms of brightness temperature. In reality, the best available values for  $T_{\rm sys}$ and $\Delta f$ also are dependent on $f$ (typically the best available instantaneous sensitivity, $T_{\rm sys}/\sqrt{\Delta f}$, is a very weak function of frequency), and this is also mitigated by, both, the fact that point source confusion is greatest at the lowest frequency, and probably more significantly, that achieving the saturation resolution requires a larger diameter telescope for lower frequencies. Nonetheless, it questions the prejudice that the best frequency at which to design an instrument to perform a blind SZ search is near the maximum in the flux density of the SZ effect at  $f\approx 150{\rm GHz}$.

In Fig.~\ref{fig:tot1a} we have plotted (\ref{ntot}) as a function of $T_{\rm lim}$ and $\theta_{\rm FWHM}$ assuming that one can tune any beam smoothing to the size of any cluster larger than the beam size, that is, $N_{\rm b}$ can take any value between $1$ and $\infty$ (that is, $\theta_{\rm FWHM}^{\rm max}=\infty$) and using $z_{\rm max}=1.5$. The computed values have been normalized so that the maximum value is one, and hence the precise value plotted quantifies the efficiency of a particular observational setup.

We see that there is a degenerate line with $T_{\rm lim}=4T_{\sigma}\approx 55\mu{\rm K}/(\theta_{\rm FWHM}/1^{\prime})$ for $\theta_{\rm FWHM}$ less than the saturation resolution of $2^{\prime}$, which can also be phrased in terms of the integrated $y$-distortion limit $Y_{\rm lim}\approx 2\times 10^{-5}/(\theta_{\rm FWHM}/1^{\prime})$. Therefore, we have shown that for sufficiently high resolution surveys, it should be possible to perform a survey which is optimal in the sense of finding the largest number of clusters for a given receiver sensitivity. For lower resolution surveys, there is still an optimum for a given value of $\theta_{\rm FWHM}^{\rm min}$, but this would be sub-optimal relative to performing a survey with the same receivers, but using a telescope with higher resolution. 

It is simple to understand the degenerate line in Fig.~\ref{fig:tot1a}: one has to achieve a given brightness temperature sensitivity at the saturation resolution of $2^{\prime}$ in order to detect the clusters with masses $\sim 10^{14}M_{\odot}$ which it is most efficient to detect. The required sensitivity at a higher resolution is that, which when smoothed, is  optimal at the saturation resolution, that is, $T_{\rm lim}\theta_{\rm FWHM}\propto S_{\rm lim}/\theta_{\rm FWHM}$ is resolution independent. This  corresponds to it being independent of $N_{\rm b}$ as discussed earlier and this is an important property of a single dish, total power measurement.

We have done the same calculation, but this time allowing for no beam smoothing and this is presented in Fig.~\ref{fig:tot1b}. The result is very different to that in Fig.~\ref{fig:tot1a}. There is no degenerate line and the optimal configuration is localized around $\theta_{\rm FWHM}\approx 2^{\prime}$. This is because smaller beams find it more difficult to detect objects which are larger than the beam, when no beam smoothing is allowed. 

\begin{figure} 
\setlength{\unitlength}{1cm}
\centerline{\psfig{file=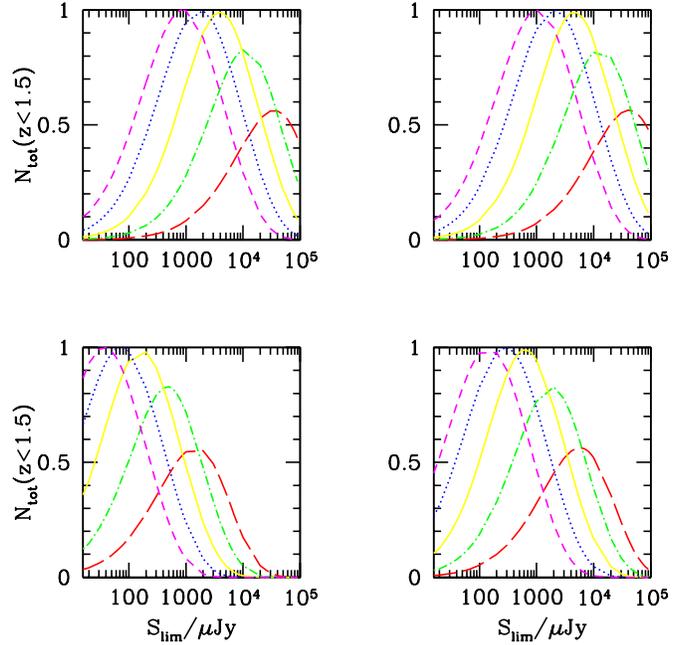,width=9cm,height=9cm}}
\caption{The relative number of clusters with $z<1.5$ that would be found in a survey with $f=15{\rm GHz}$ (bottom left), $f={\rm 30}{\rm GHz}$ (bottom right), $f=90{\rm GHz}$ (top left) and $f=150{\rm GHz}$ (top right). In each case $\theta_{\rm FWHM}=0.5^{\prime}$ (short dashed line), $1^{\prime}$ (dotted line), $2^{\prime}$ (solid line), $4^{\prime}$ (dashed-dotted line) and $8^{\prime}$ (long dashed line). The optimal survey strategy would be to choose $S_{\rm lim}^{\rm min}$ to take the value where the curves are maximum. They are normalized by the largest possible number of clusters that could be found in survey with that frequency and hence we see that surveys with $\theta_{\rm FWHM}^{\rm min}>2^{\prime}$ are not optimal. The curves represent a slice through the contour plot in Fig.~\ref{fig:tot1a} expressed in flux density units which naturally introduces a dependence on $f$.}
\label{fig:ntot2}
\end{figure}

We have also computed the optimal configuration in terms of the flux density limit $S_{\rm lim}$ which will, in contrast to the non-standard temperature units as defined above, be dependent on the observation frequency. This is plotted in Fig.~\ref{fig:ntot2} for $f=15\,{\rm GHz}$, $30\,{\rm GHz}$, $90\,{\rm GHz}$ and $150\,{\rm GHz}$ and a range of different values of $\theta_{\rm FWHM}^{\rm min}$. The different values of $\theta_{\rm FWHM}$ correspond to slices of constant $\theta_{\rm FWHM}$ through the contour plot in Fig.~\ref{fig:tot1a}. We see that in each case there is a maximum in the curve at a particular value of $S_{\rm lim}$ and these are tabulated in Table \ref{tab:surveys} for various possible configurations. Note that once again those with $\theta_{\rm FWHM}>2^{\prime}$ are seen to be non-optimal. The corresponding thermodynamic temperature limits are presented in Table \ref{tab:ttt}.

The existence of an optimum at $S=S_{\rm opt}$ can be understood by considering $N(>S)\propto S^{-\alpha(S)}$, the number of clusters detected with a flux limit per beam greater than $S$. It is clear that for $S\ll S_{\rm opt}$ the function $\alpha(S)<2$ due to the properties of the mass function and for $S\gg S_{\rm opt}$ we have that $\alpha>2$ since the detection of objects very much larger than the beam is very inefficient. $S_{\rm opt}$ is defined by $\alpha(S_{\rm opt})=2$ and it is clear that it will increase with the  beamsize.

\begin{table}
\centering
\begin{tabular}{ccccc}
\hline
$\theta_{\rm FWHM}^{\rm min}$ & $f=15{\rm GHz}$ & $30{\rm GHz}$ & $90{\rm GHz}$ & $150{\rm GHz}$ \\
\hline 
\hline 
$0.5^{\prime}$ & 35 & 140 & 870 & 1030 \\
$1^{\prime}$ & 70 & 280 & 1730 & 2030 \\
$2^{\prime}$ & 160 & 620 & 3830 & 4500 \\
$4^{\prime}$ & 480 & 1870 & 11520 & 13470 \\
$8^{\prime}$ & 1500 & 5810 & 35690 & 41620 \\
\hline 
\hline 
\end{tabular}
\caption{The optimum limiting flux densities (in $\mu{\rm Jy}$) for different set ups, which correspond to the maxima of the graphs curves in Fig.~\ref{fig:ntot2}. We have assumed the idealized total power setup allowing for beam smoothing  described in section~\ref{total}. The numbers are quoted to the nearest $10\mu{\rm Jy}$ except for $\theta_{\rm FWHM}=0.5^{\prime}$ and $f=15{\rm GHz}$ where the flux is quoted to the nearest $5\mu{\rm Jy}$.}
\label{tab:surveys}
\end{table}

\begin{table}
\centering
\begin{tabular}{ccccc}
\hline
$\theta_{\rm FWHM}^{\rm min}$ & $f=15{\rm GHz}$ & $30{\rm GHz}$ & $90{\rm GHz}$ & $150{\rm GHz}$ \\
\hline 
\hline 
$0.5^{\prime}$ & 230 & 225 & 185 & 110 \\
$1^{\prime}$ & 110 & 110 & 90 & 55 \\
$2^{\prime}$ & 60 & 60 & 50 & 30\\
$4^{\prime}$ & 45 & 45 & 40 & 25 \\
$8^{\prime}$ & 35 & 35 & 30 & 20 \\
\hline 
\hline 
\end{tabular}
\caption{The thermodynamic temperature limits (in $\mu{\rm K}$) quoted which correspond to the flux density limits presented in Table \ref{tab:surveys} to the nearest $5\mu{\rm K}$. Note that these can also be computed approximately from $T_{\rm lim}^{\rm THERM}\approx 55\mu{\rm K}|g(x)|/(\theta_{\rm FWHM}/1^{\prime})$ for $\theta_{\rm FWHM}\le 2^{\prime}$.}
\label{tab:ttt}
\end{table}

\section{More realistic observations}
\label{real}

\subsection{Systematics in total power measurements}

The ability to perform beam smoothing is not a given, but we have shown that it is an essential feature of an SZ survey instrument with resolution $<2^{\prime}$. It appears that a number of ground based single dish experiments have been able to make accurate measurements of the angular power spectrum of the primary CMB anisotropies over a wide range of scales. However, the reliable detection of extended objects like clusters requires the maps to not only have the power distributed correctly, but also requires the phases to be accurate.

\begin{figure} 
\setlength{\unitlength}{1cm}
%\vskip -2cm
\centerline{\psfig{file=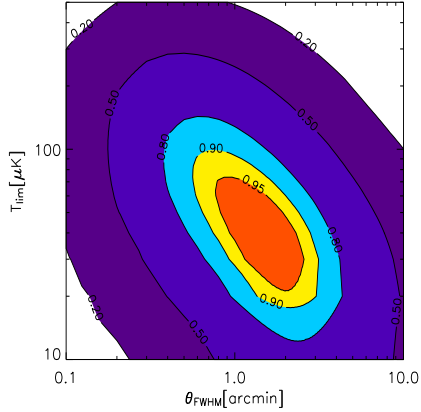,width=6.0cm,height=6.0cm}}
%\vskip -1cm
\caption{Equivalent to Fig.~\ref{fig:tot1b}, but this time with $\theta_{\rm FWHM}^{\rm max}=3\theta_{\rm FWHM}^{\rm min}$. There is a degenerate line in the same direction as in Fig.~\ref{fig:tot1a}, but the finite amount of allowed beam smoothing prevents very high resolution instruments being efficient for this purpose.}
\label{fig:tot3a}
\end{figure}

\begin{figure} 
\setlength{\unitlength}{1cm}
%\vskip -2cm
\centerline{\psfig{file=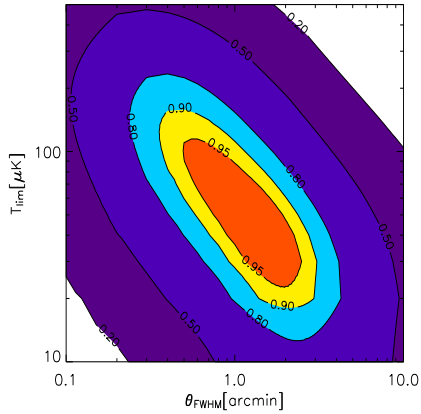,width=6.0cm,height=6.0cm}}
%\vskip -1cm
\caption{Equivalent to Fig.~\ref{fig:tot3a}, but this time with $\theta_{\rm FWHM}^{\rm max}=5\theta_{\rm FWHM}^{\rm min}$. Notice that the degenerate line extends to smaller values of $\theta_{\rm FWHM}$ than in Fig.~\ref{fig:tot3a}, but that is does not extended to zero as in Fig.~\ref{fig:tot1b}.}
\label{fig:tot3b}
\end{figure}

A variety of instrumental systematics, such as baseline offsets caused by the receivers and the atmosphere, could lead to artifacts which might destroy the integrity of a map once it is smoothed to low resolution. In terms of the design of an SZ survey this leads to the possibility that $\theta_{\rm FWHM}^{\rm max}<\infty$. In what follows we will assume that some beam smoothing is possible, but that this is limited to some constant multiple of $\theta_{\rm FWHM}^{\rm min}$. In particular, we consider the effects of $\theta_{\rm FWHM}^{\rm max}=3\theta_{\rm FWHM}^{\rm min}$ and $\theta_{\rm FWHM}^{\rm max}=5\theta_{\rm FWHM}^{\rm min}$ in Figs.~\ref{fig:tot3a} and \ref{fig:tot3b}. We see that in both cases the degenerate line seen in Fig.~\ref{fig:tot1a} is truncated at small values of $\theta_{\rm FWHM}$ (higher values for $\theta_{\rm FWHM}^{\rm max}=3\theta_{\rm FWHM}^{\rm min}$). This can be understood by realizing that instruments with very small value of $\theta_{\rm FWHM}$ are unable to achieve critical value of $2^{\prime}$ by beam smoothing when these constraints are placed on $\theta_{\rm FWHM}^{\rm min}$. We conclude from this, that although extra resolution might be desirable (for example, to allow  subselection within the sample), excessive amounts can be damaging to an instrument's survey potential.

\subsection{Beam switching}
\label{real2}

Beam switching can be used to subtract out the effects of the atmosphere in single dish observations when it varies on timescales shorter than the characteristic integration time. To investigate its effect, we will consider two Gaussian beams separated by angle $\theta_{\rm s}$ on the sky. It is likely that the properties of  the atmosphere will prevent $\theta_{\rm s}$ from being greater than some value (probably somewhere between $5^{\prime}$ and $10^{\prime}$), but from the point of view of our discussion we will assume that it can take any value between $\theta_{\rm FWHM}^{\rm min}$ and $\infty$. 
 
First, let us consider a flat sky approximation with coordinates $x$ and $y$, and let $\tau_{\rm e}(x,y)$ be the optical depth to the SZ effect along the line of sight parameterized by $(x,y)$. If a Gaussian beam, with width denoted by $\sigma$, is centred at the point $(x_{\rm s},0)$ then the resultant flux will be proportional to 
\be
\int dx\int dy \,\tau_{\rm e}(x,y)\exp\left[-{1\over 2 \sigma^2}\left\{(x-x_{\rm s})^2+y^2\right\}\right]\,,
\ee 
where the range of integration is $x^2+y^2<R^2_{\rm vir}$. The optical depth $\tau_{\rm e}(x,y)=\tau_{\rm e}(0)\zeta(r)$ is only dependent on the radial coordinate $r=\sqrt{x^2+y^2}$. By making the coordinate change $x=r\cos\phi$ and $y=r\sin\phi$, one can write this integral as 
\be
\tau_{\rm e}(0)\int_0^{R_{\rm vir}}\,dr\,r\zeta(r)\exp\left[-{r^2\over 2\sigma^2}\right]K(r,x_{\rm s})\,,
\label{kk1}
\ee
where
\be 
K(r,x_{\rm s})=\exp\left[-{x_{\rm s}^2\over 2\sigma^2}\right]\int_0^{2\pi}\,d\phi \exp\left[{rx_{\rm s}\over\sigma^2}\cos\phi\right]\,.
\label{kk}
\ee
If $K(r,x_{\rm s})=1$ then this expression is that of a Gaussian beam observing the centre of the object as used in our discussion of idealized total power measurements. This factor represents the effects of a non-centred beam.
 
One can re-write (\ref{kk}) in terms of the modified Bessel function $I_0(z)$ and, by considering the difference between a beam pointing at the origin and one pointing at $x=x_{\rm s}$, an effective beam $B_{\rm eff}(\theta)$ can be constructed which represents the effects of a beam switch. If we also make the transition from the flat-sky coordinate $r$ to the angular coordinate $\theta$, factor ${\cal I}(B,\zeta)$ becomes 
\be 
{\cal I}(B_{\rm eff},\zeta)={\int\,d\Omega\,B_{\rm eff}(\theta)\zeta(\theta)\over \int\, d\Omega\,\zeta(\theta)}\,,
\ee
where 
\be 
B_{\rm eff}(\theta)=\exp\left(-{\theta^2\over 2\sigma_{\rm b}^2}\right)
\left[1-\exp\left(-{\theta_{\rm s}^2\over 2\sigma_{\rm b}^2}\right)I_0\left(
{\theta\theta_{\rm s}\over \sigma_{\rm b}^2}\right)\right]\,.
\label{switch}
\ee
In Appendix B, we show that, under sensible assumptions as the form of $\zeta(\theta)$, ${\cal I}(B_{\rm eff},\zeta)\rightarrow 0$ as $\theta_{\rm vir}/\theta_{\rm FWHM}\rightarrow\infty$, in which case it is clear that $M_{\rm lim}\rightarrow\infty$ as $z\rightarrow 0$ in contrast to the case of a Gaussian beam profile for which $M_{\rm lim}$ tends to a finite value. The reason for this is simple: as one approaches $z=0$ the clusters become much larger, but the beam switching makes it impossible to find objects which are very much larger than the switching angle. This phenomenon will play a key role in our subsequent discussion.

\begin{figure} 
\setlength{\unitlength}{1cm}
\centerline{\psfig{file=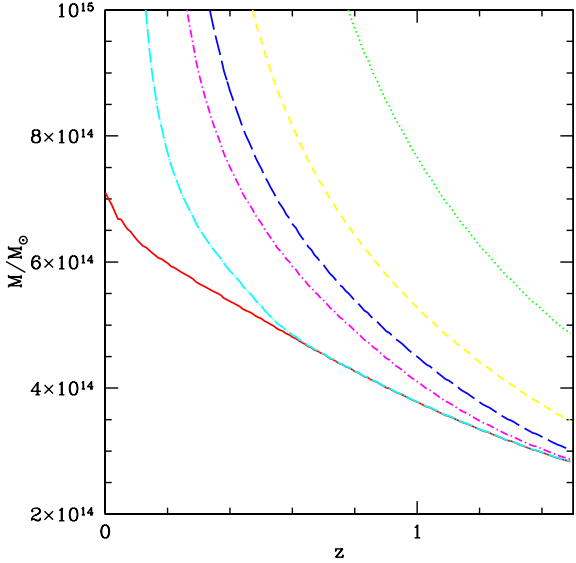,width=8cm,height=8cm}}
\caption{The solid line is the limiting mass of a survey with $f=30{\rm GHz}$, $\theta_{\rm FWHM}^{\min}=1^{\prime}$ and $S_{\rm lim}^{\rm min}=500\mu{\rm Jy}$ as for that in Fig.~\ref{fig:mlim1}. The other curves are for different values of $\theta_{\rm s}$. The dotted line is $\theta_{\rm s}=2^{\prime}$, the short dashed line is $3^{\prime}$, the long dashed line is $4^{\prime}$, the short dashed-dotted line is $5^{\prime}$ and the long dashed-dotted line is $10^{\prime}$. We see that the effect of beam switching is to make the survey less efficient particularly at low redshifts where clusters have a much larger angular diameter. None of the curves presented here have included any beam smoothing.}
\label{fig:mlim_sw1}
\end{figure}

We have computed $M_{\rm lim}(z)$ for $\theta_{\rm FWHM}=1^{\prime}$, $f=30{\rm GHz}$ and $S^{\rm min}_{\rm lim}=500\mu{\rm Jy}$ for a range of values for $\theta_{\rm s}$ compared to that for no switching and the results are presented in Fig.~\ref{fig:mlim_sw1}. At this stage, we ignore the effects of beam smoothing. We see that for large values of $\theta_{\rm s}\approx 10^{\prime}$, the limiting mass is the same as for no switching when $z>0.5$. However, as suggested above, we see that for all finite values of $\theta_{\rm s}$, $M_{\rm lim}\rightarrow\infty$ as $z\rightarrow 0$. As $\theta_{\rm s}$ is reduced the value of z above which the no switching case is a good approximation increases. For $\theta_{\rm s}\approx 3^{\prime}$ the limiting mass is affected ($\sim 50\%$) even at $z\approx 1$; given that the number of clusters found in the survey is very sensitive to the limiting mass this can lead to the number of clusters detected being significantly reduced.

\begin{figure} 
\setlength{\unitlength}{1cm}
\centerline{\psfig{file=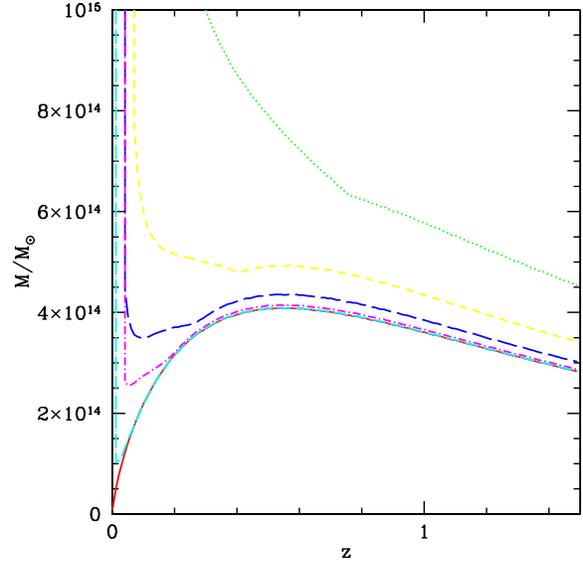,width=8cm,height=8cm}}
\caption{This  illustrates the possible effects of beam smoothing when beam switching is involved. The curves represent the same values of $\theta_{\rm s}$ as in Fig.~\ref{fig:mlim_sw1} and the parameters of the survey are the same. The solid line is that for no switching and is equivalent to the solid line in Fig.~\ref{fig:mlim3}. We allow beam smoothing up to the switching angle; that is, we set $\theta_{\rm FWHM}^{\rm max}=\theta_{\rm s}$. We see that in each case the limiting mass tends to $\infty$ as $z\rightarrow 0$ (see Appendix B). For $\theta_{\rm s}>5^{\prime}$ the limiting mass is almost equivalent to that for no switching except for $z<0.1$ where, due to the small comoving volume, there are very few clusters available to be detected.}
\label{fig:mlim_sw2}
\end{figure}

The calculation discussed above does not take into account beam smoothing which is more subtle than in the case of a single-dish observation. In the case of switching it will be impossible to perform beam smoothing beyond the switching angle, that is, $\theta_{\rm FWHM}^{\rm max}=\theta_{\rm s}$. We illustrate the effects of this on the limiting mass the same values of $f$, $\theta_{\rm FWHM}^{\rm min}$ for a range of values of $\theta_{\rm s}$ in Fig.~\ref{fig:mlim_sw2} compared to the case of no switching. Once again we see that for all finite values of $\theta_{\rm s}$,  $M_{\rm lim}\rightarrow\infty$ as $z\rightarrow 0$, but that for $\theta_{\rm s}>4^{\prime}$ the limiting mass is only significantly different when compared to the case of no switching when $z<0.2$. Since there are only very few clusters to be detected at these low redshifts, we conclude that for $\theta_{\rm s}>4^{\prime}$ and $\theta_{\rm FWHM}^{\rm min}=1^{\prime}$ beam switching does not affect the number of clusters detected in a given survey.
    
\begin{figure} 
\setlength{\unitlength}{1cm}
\centerline{\psfig{file=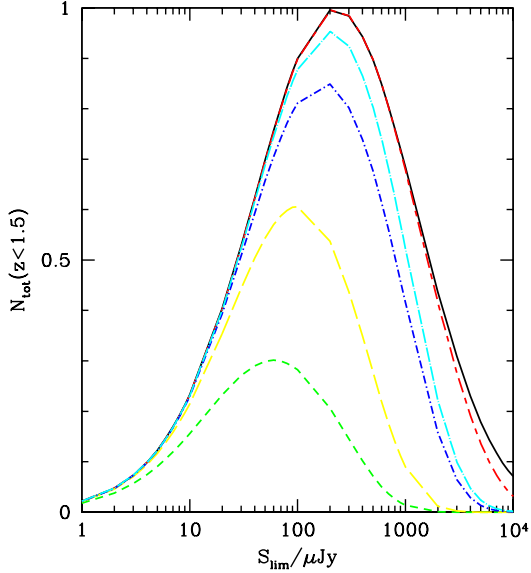,width=8cm,height=8cm}}
\caption{The relative number of clusters with $z<1.5$ that would be found in a beam switch surveys with $f=30{\rm GHz}$ and $\theta_{\rm FWHM}=1^{\prime}$ for different values of the switching angle $\theta_{\rm s}$. The curves are labelled as in Figs.~\ref{fig:mlim_sw1} and \ref{fig:mlim_sw2}. The solid line is that for no switching and the curves are normalized to one at the maximum for this at around $S_{\rm lim}\approx 280\mu{\rm Jy}$. As the switching angle decreases,  searching for clusters becomes less efficient, and the optimal limiting flux decreases to around $S_{\rm lim}\approx 100\mu{\rm Jy}$ for $\theta_{\rm s}=3^{\prime}$ which is about 60\% efficient.}
\label{fig:ntot_sw}
\end{figure}

Now that we have a method for computing the limiting mass in the case of beam switching, we can attempt to investigate how to make the optimum choice for the survey depth as we did for the single dish, total power case. We will assume that one can beam smooth up to $\theta_{\rm FWHM}^{\rm max}=\theta_{\rm s}$ and will do this only for the case of $\theta_{\rm FWHM}^{\rm min}=1^{\prime}$ and $f=30{\rm GHz}$, and will work in terms of $S_{\rm lim}$ (this can be converted into $T_{\rm lim}$ using (\ref{st1})). Fig.~\ref{fig:ntot_sw} is the equivalent to Fig.~\ref{fig:ntot2}, but for a range of values of $\theta_{\rm s}$. We see that, so long as the optimal choice of $S_{\rm lim}$ is made, the number of objects is only marginally reduced for $\theta_{\rm s}>5^{\prime}$. For $\theta_{\rm s}=4^{\prime}$ the number of objects found is only reduced to about $80\%$ at the optimum, while for $\theta_{\rm s}=3^{\prime}$ one can only achieve about $60\%$ of that possible for no switching. The number of objects found is significantly reduced for lower values of $\theta_{\rm s}$. Moreover, we see that the optimal value of $S_{\rm lim}$ is reduced from $280\mu{\rm Jy}$ (see Table 1) for no switching to around $200\mu{\rm Jy}$ ($T_{\rm lim}^{\rm THERM}\approx 80\mu{\rm K}$) for $\theta_{\rm s}=4^{\prime}$ and $100\mu{\rm Jy}$ ($T_{\rm lim}\approx 40\mu {\rm K}$)  for $\theta_{\rm s}=3^{\prime}$. Although the optimal value of $S_{\rm lim}$ only changes by about a factor 3 in going from $\theta_{\rm s}=\infty$ to $\theta_{\rm s}=3^{\prime}$, the number of objects found for larger values of $S_{\rm lim}$ falls off much more quickly than in the no switching case. Both these effects can be explained by the fact that it is difficult to detect objects at very low redshifts for small values of $\theta_{\rm s}$: as $\theta_{\rm s}$ decreases one has to make surveys which are deeper and deeper in order to find anything, and very shallow surveys are unlikely to find anything at all since they are only sensitive to the largest, very nearby clusters which are effectively resolved out by the beam switch.

\subsection{Aspects of interferometric observations}
\label{real1}

As we have already pointed, out making total power, single-dish measurements are not always possible. Interferometric methods using a telescope array provide an alternative to this, which can reduce the sensitivity of the measurements  to a range of systematic effects notably those due to the atmosphere. The techniques involved (interferometers work in Fourier space, whereas single dishes work in real space) are very different. In this section we will discuss how, under certain circumstances, they can be thought of as being a special case of the total power measurements which we described in section \ref{total} and we will outline how they can be incorporated into our framework. There are definitely some caveats to their application to an interferometer which we will ignore, for the most part, here.

Let us consider an  array with $n_{\rm r}$ receivers mounted on telescopes each with diameter $D$,  shortest baseline $s\ge D$ and longest baseline $d$ as measured between the centres of the dishes which are furthest apart. Such an array will have $B=n_{\rm r}(n_{\rm r}-1)/2$ baselines and can be thought of as having a synthesized beam which can be represented by a Gaussian with FWHM $\theta_{\rm FWHM}^{\rm synth}\approx\lambda/d$ and a primary beam, which corresponds to the field-of-view, which is a Gaussian with FWHM $\theta_{\rm FWHM}^{\rm prim}\approx\lambda/D$. The shortest baseline sets the scale of the largest structure which can be observed. It is our claim that in the language of section 4, we can approximate the effects of an interferometer by a Gaussian beam with $\theta_{\rm FWHM}^{\rm min}\approx\lambda/d$ and a maximum beam angle of $\theta_{\rm FWHM}^{\rm max}\approx\lambda/s$ assuming that the density of visibilities is uniform in the $u$-$v$ plane, but with two important differences. We note that the assumption that the beam is approximately Gaussian requires that the number of baselines is sufficiently large so that the negative response inevitable in interferometric observations is negligible.

The first important difference is in terms of the sensitivity of the array. Such an arrangement will have a flux sensitivity corresponding to a single dish with the same overall collecting area $\propto n_{\rm r}D^2$, but the temperature sensitivity is diluted by the so-called `filling factor' $\eta_{\rm F}=n_{\rm r}D^2/(D+d)^2\le 1$ ($\approx n_{\rm r}D^2/d^2$ if $d\gg D$), such that 
\be
T_{\sigma}={T_{\rm sys}\over \eta_{\rm F}\eta\sqrt{t\Delta f}}\,,
\ee
from which we can see that any survey with a fixed integration time should satisfy 
\be
{T_\sigma\over\sqrt{\Delta\Omega}}=\sqrt{4\log 2\over\pi}{T_{\rm sys}\over\theta^{\rm synth}_{\rm FWHM}\eta\sqrt{\eta_{\rm F} n_{\rm r}t\Delta f}}\,,
\ee
and hence the corresponding instrumental figure of merit is given by
\be 
{\cal M}_{\rm INT}={\eta_{\rm F}n_{\rm r}\eta^2\Delta f\over T^2_{\rm sys}}={\eta_{\rm F}n_{\rm r}\eta^2\Delta f\over (T_{\rm sys}^{\rm THERM})^2}[g(x)]^2\,.
\ee
The limit $\eta_{\rm F}=1$ corresponds to a fully-filled array and is equivalent in some sense to a single dish with focal plane array of $n_{\rm b}=n_{\rm r}$ elements. We see that an interferometer which is fully-filled will be the most efficient. This is not particularly important in the context of our discussion here since it just means that it will take longer to achieve the same temperature sensitivity, or given a fixed integration time, one would be able to cover less area to the same depth.

The other important difference which needs to be taken into account can be  very significant in terms of the optimum depth of a survey. The quantity $T_{\rm lim}\theta_{\rm FWHM}$ need not be independent of resolution for an interferometer, as it is for a single dish; this is a consequence of the density of visibilities within the $u$-$v$ plane generated by positions of the antennae in the interferometer. We will describe an interferometer for which  $T_{\rm lim}\theta_{\rm FWHM}$ is independent of resolution as being uniformly filled. 

We can attempt to model an interferometer within our framework by assuming that the baseline lengths can be grouped into $m$ bins and that there is an approximate resolution $\theta_{\rm FWHM}^{i}\approx \lambda/b_{i}$ corresponding to each of the bins $1\le i\le m$, where $b_{i}$ is the average length of the baselines in bin $i$. There should be no significant error introduced by binning of the baselines if $m$ sufficiently large. We will assume that at each resolution the response of the interferometer can be modelled by a Gaussian beam with FWHM $\theta_{\rm FWHM}^{i}$ which should be a good approximation so long as the number of baselines with length $\le b_i$ is sufficiently large. By reconstructing the beam in this way, we are effectively ignoring the data from long baselines. The flux limit corresponding to resolution $\theta_{\rm FWHM}^{i}$ will be given by 
\be 
S_{\rm lim}^{i}=\sqrt{B\over B^{i}}S_{\rm lim}^{\rm min}\,,
\ee
where $B^{i}$ is the number in bins with length $\le b_{\rm i}$. 

 There is no particular reason, a priori, why $T_{\rm lim}^{i}\theta_{\rm FWHM}^{i}\propto S_{\rm lim}^{i}/\theta^{i}_{\rm FWHM}$ should be independent of $i$, which is what would be required for the interferometer to be uniformly filled! Given a specific set of baseline lengths, it would be simple to compute the overall limiting mass as that which corresponds to the envelope of the individual values of $i$.

One can re-phrase this discussion in terms of $u$-$v$ plane of the interferometer. If $u=|{\bf u}|$ is the amplitude of the Fourier variable ${\bf u}=(u\sin\phi,u\cos\phi)$, defined in multiples of the wavelength $\lambda$, then $\theta_{\rm FWHM}^{\rm max}=1/u_{\rm min}$ and $\theta_{\rm FWHM}^{\rm min}=1/u_{\rm max}$, where $u_{\rm max}$ and $u_{\rm min}$ are the maximum and minimum values of $u$. In principle, one can create a beam, which we will assume is Gaussian, corresponding to any value $u_{\rm min}<u<u_{\rm max}$, with FWHM and flux density limit given by 
\be 
\theta_{\rm FWHM}^u={u_{\rm max}\over u}\theta_{\rm FWHM}^{\rm min}\,,\qquad S_{\rm lim}^{u}=\left({B^{u_{\rm max}}\over B^{u}}\right)^{1/2}S_{\rm lim}^{\rm min}\,,
\ee
where 
\be
B^{u}=2\pi\int_{\rm u_{\rm min}}^{u} u^{\prime}\rho(u^{\prime})du^{\prime}\,,
\ee
is the number of visibilities $<u$ and $\rho(u)$ is the density of visibilities. The case of a uniformly filled interferometer can then be seen to correspond to $\rho(u)=\rho_0$, that is, a uniform density of visibilities in the $u$-$v$ plane.

An interesting parameterization of interferometers in this context is to consider $\rho(u)\propto u^{\gamma}$ where $\gamma\ge 0$. For $u\gg u_{\rm min}$, $B^u\propto u^{\gamma+2}$ and hence
\be
S^{u}_{\rm lim}=\left({u_{\rm max}\over u}\right)^{1+{1\over 2}\gamma}S^{\rm min}_{\rm lim}\,.
\ee
If we choose $N_{\rm b}=(u_{\rm max}/u)^{1/2}$ in order detect clusters which cover $N_{\rm b}$ beams, then we find that
\be
S^{u}_{\rm lim}=N_{\rm b}^{{1\over 2}+{1\over 4}\gamma}S_{\rm lim}^{\rm min}\,.
\ee
We will explore this parameterization in future work, but it is clear that unless $\gamma=0$ the behaviour of an interferometer when smoothing will differ from that of the idealized single dish case discussed in the previous section.

\begin{figure} 
\setlength{\unitlength}{1cm}
%\vskip -2cm
\centerline{\psfig{file=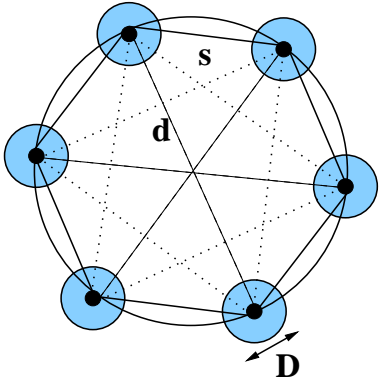,width=5.0cm,height=5.0cm}}
%\vskip -1cm
\caption{A realization of the  simple interferometer arrangement discussed in the test with $n_{\rm r}=6$. $d$ and $s$ is the longest and shortest baselines respectively. $D$ is the dish size. In this case there are three different types of baseline which, in order of increasing length have solid, dotted and dashed lines.}
\label{fig:int}
\end{figure}

In order to illustrate this further,  we will consider a very specific arrangement of the telescopes. Let us assume that they  are arranged in a circular configuration. In order to make our discussion simpler we will assume that $n_{\rm r}$ is even, in which case there $n_{\rm r}/2$ different baseline lengths.  If these baseline lengths are denoted $b_i$ for $1\le i\le n_{\rm r}/2$ then 
\be 
b_{i}=d\sin\left[{i\pi\over n_{\rm r}}\right]\,,
\ee
where there are $n_{\rm r}$ of length $b_{\rm i}$ for  $i\le n_{\rm r}/2-1$ and $n_{\rm r}/2$ with length $b_{n_{\rm r}/2}=d$. A diagram illustrating the arrangement for $n_{\rm r}=6$ is presented in Fig.~\ref{fig:int}.

The shortest baseline is given by $s=d\sin[\pi/n_{\rm r}]$ and hence the ratio of the shortest to longest baseline, which is equal to $\theta_{\rm FWHM}^{\rm min}/\theta_{\rm FWHM}^{\rm max}$ is given by 
\be 
{s\over d}={\theta_{\rm FWHM}^{\rm min}\over \theta_{\rm FWHM}^{\rm max}}=\sin\left[{\pi\over n_{\rm r}}\right]\,.
\ee
The largest possible filling factor achievable with such a configuration is when $s=D$, that is, when all the dishes in the circular arrangement are touching. In this case $\eta_{\rm F}=n_{\rm r}/(1+{\rm cosec}(\pi/n_{\rm r}))^2$ which tends to zero like $1/n_{\rm r}$ for large $n_{\rm r}$.

A range of resolutions are possible between the minimum and maximum , given by
\be 
\theta_{\rm FWHM}^{i}={\lambda\over b_{\rm i}}={\rm cosec}\left[{i\pi\over
n_{\rm r}}\right]\theta_{\rm FWHM}^{\rm min}\,,
\ee
which result in the limiting flux density  of
\be 
S_{\rm lim}^{i}=\sqrt{n_{\rm r}-1\over 2i}S_{\rm lim}^{\rm min}\,, 
\ee
for $i\le n_{\rm r}/2-1$ and limiting temperature of 
\be 
T_{\rm lim}^{i}=\sqrt{n_{\rm r}-1\over 2i}\sin^2\left[{i\pi\over n_{\rm r}}\right]T_{\rm lim}^{\rm min}\,.
\ee

\begin{figure} 
\setlength{\unitlength}{1cm}
%\vskip -2cm
\centerline{\psfig{file=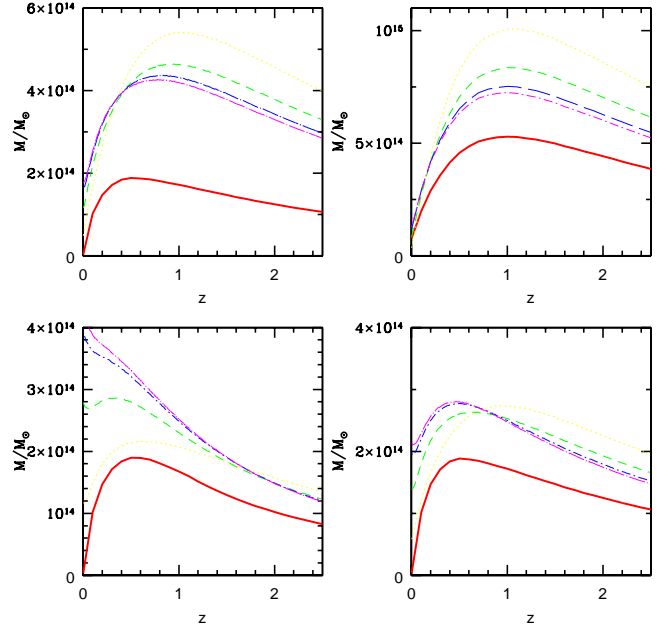,width=9.0cm,height=9.0cm}}
%\vskip -1cm
\caption{The mass limits for the simple interferometric configuration discussed int the text with $n_{\rm r}=8$. The bottom left has $\theta_{\rm FWHM}^{\rm min}=1^{\prime}$ and $S_{\rm lim}^{\rm min}=300\mu{\rm Jy}$, the bottom right $\theta_{\rm FWHM}^{\rm min}=2^{\prime}$ and $S_{\rm lim}^{\rm min}=600\mu{\rm Jy}$, the top left $\theta_{\rm FWHM}^{\rm min}=4^{\prime}$ and $S_{\rm lim}^{\rm min}=2{\rm mJy}$, and the top right has $\theta_{\rm FWHM}^{\rm min}=8^{\prime}$ and $S_{\rm lim}^{\rm min}=5.8{\rm mJy}$. In each case the solid line corresponds to the equivalent single dish observation assuming the maximum filling factor allowed for $n_{\rm r}=8$, $\eta_{\rm F}=0.6$, the dotted line to $i=1$, the short dashed line to $i=2$, the long dashed line for $i=3$, and the dot-dashed line  to $i=4$. The overall mass limit which would be achieved for each of the interferometers would be the envelope of the curves corresponding to $i=1$ to 4. Note that in each case the mass limit for the  equivalent single dish is better than envelope of those for the corresponding to the interferometer.}
\label{fig:mlim_int}
\end{figure}

It is clear that this interferometer is not uniformly filled for finite values of $n_{\rm r}$ since 
\be
T_{\rm lim}^{i}\theta_{\rm FWHM}^{i}=\sqrt{n_{\rm r}-1\over 2i}\sin\left[{i\pi\over n_{\rm r}}\right]T_{\rm lim}^{\rm min}\theta_{\rm FWHM}^{\rm min}\,,
\ee
which is clearly not independent of $i$. Such a configuration is close to being uniformly filled for the longest baselines if $n_{\rm r}$ is large. If we define $j=n_{\rm r}/2-i$ and assume that $n_{\rm r}$ is large then 
\be 
T_{\rm lim}^{i}\theta_{\rm FWHM}^{i}\approx T_{\rm lim}^{\rm min}\theta_{\rm FWHM}^{\rm min}+{\cal O}(j)\,,
\ee
that is, the longest baselines are close to being uniformly filled. But even in this case, the filling factor is not uniform for small $i$ where the discreteness takes hold.

We have plotted the mass limits for $1\le i\le 4$ for $n_{\rm r}=8$ in Fig.~\ref{fig:mlim_int} for a range of values of $\theta_{\rm FWHM}^{\rm min}$ along with the predictions of a single dish with the same maximum resolution. In this case the best filling factor possible is $\eta_{\rm F}=8/(1+{\rm cosec}(\pi/8))^2\approx 0.6$. The actual mass limit that one would achieve with an interferometer is given by the envelope of the curves for $i=1$ to 4, and it is clear to see that one always does worse in terms of detecting clusters with an interferometer when compared to an ideal single dish with the same resolution. We note that despite this advantage, systematics are nearly always worse for single dish observations.

The values of $S_{\rm lim}^{\rm min}$ have, in each case, been chosen to be the values which are close to the optimal survey strategy in the case where the interferometer is uniformly filled (but this might not be optimal here). There are some interesting aspects of the problem which are raised by Fig.~\ref{fig:mlim_int}. We see that for $\theta_{\rm FWHM}^{\rm min}=1^{\prime}$ if $z<1$ the mass limit is ostensibly defined by that for $i=1$ corresponding to the shortest baselines, whereas for $\theta_{\rm FWHM}=8^{\prime}$ it is that for $i=4$ corresponding to the longest baselines which dominates except for the very smallest values of $z$.

\section{Are our predictions robust?}
\label{robust}

\subsection{Cosmological parameters}

In making our calculations we found it necessary to make choice of cosmological parameters. It is unlikely that the total number of clusters found in a given survey is very strongly dependent on the parameters $h$, $w_0$ and $w_1$ although how they are distributed as a function of $z$ should be. However, it is known that it will be strongly dependent on $\sigma_8$ and $\Omega_{\rm m}$. We have made sensible choices for these parameters in the light of presently available cosmological observations, but it is perfectly possible that our choices are incorrect. The value we have used, $\sigma_8=0.9$, is that preferred by measurements of the primary anisotropies in the CMB and weak lensing. However, measurements of the X-ray luminosity function of clusters and the gas fraction therein suggest that $\sigma_8\approx 0.7$~\cite{ASF}.

In light of this discrepancy, we have performed our calculations for a range of different values of $\sigma_8$ between 0.7 and 1.1. We find that there is a degenerate line in the equivalent of Fig.~\ref{fig:tot1a} for every value of $\sigma_8$ we tried  and that a survey can be optimal for $\theta_{\rm FWHM}<2^{\prime}$. However, the normalizing amplitude is modified by $\sigma_8$. We find that 
\be
T_{\rm lim}=55\mu{\rm K}\left({\sigma_8\over 0.9}\right)^{4.1}\,,
\ee
at $\theta_{\rm FWHM}=1^{\prime}$. The strong dependence on $\sigma_8$ is due to the exponential dependence of the number of clusters on it.

\begin{figure} 
\setlength{\unitlength}{1cm}
%\vskip -2cm
\centerline{\psfig{file=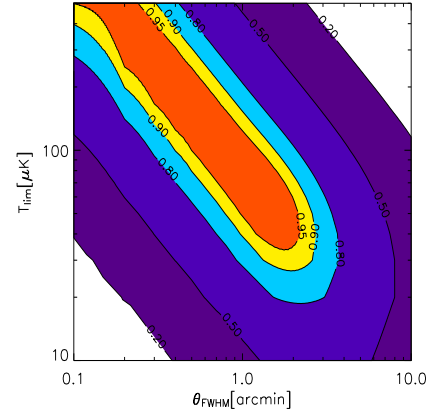,width=6.0cm,height=6.0cm}}
%\vskip -1cm
\caption{Equivalent to Fig.~\ref{fig:tot1a} for $\Omega_{\rm m}=0.5$. There is the characteristic degenerate line, but with an amplitude of $T_{\rm lim}\approx 80\mu{\rm K}$ at $\theta_{\rm FWHM}=1^{\prime}$.}
\label{fig:om}
\end{figure}

We have also investigated the effects of variations in $\Omega_{\rm m}$. Fig.~\ref{fig:om} presents the equivalent of Fig.~\ref{fig:tot1a} for $\Omega_{\rm m}=0.5$. Yet again there is a degenerate line for $\theta_{\rm FWHM}<2^{\prime}$ and for this case the amplitude for $\theta_{\rm FWHM}=1^{\prime}$ is $T_{\rm lim}\approx 80\mu{\rm K}$.

\subsection{Cluster model parameters}

Another possible uncertainty of our calculation is the cluster model which we have used. The two important features of the model in this respect are the $M$-$T$ relation (\ref{mt}) and the cluster profile function.

We have investigated the effects of varying the parameter $T_{*}$, although we note that it is perfectly possible that the relation between $M$ and $\langle T_{\rm e}\rangle_{\rm n}$ as a function of $z$ is not precisely of the form (\ref{mt}). In this case we find that the normalizing amplitude at $\theta_{\rm FWHM}=1^{\prime}$ is given by 
\be
T_{\rm lim}=55\mu{\rm K}\left({T_*\over 1.6}\right)\,,
\ee
with the degenerate line and saturation value of $\theta_{\rm FWHM}$ as before. It is easy to understand this linear dependence of $T_*$ since modifying it just leads to a corresponding modification of the temperature of each cluster. It is clear that modification to the relationship between $M$ and $\langle T_{\rm e}\rangle$ and to the $z$ dependence can led to more complicated modifications to the optimal value of $T_{\rm lim}$ dependent on how this affects the overall distribution of clusters. Once more information is available on this it might be worth investigating this further.

\begin{figure} 
\setlength{\unitlength}{1cm}
%\vskip -2cm
\centerline{\psfig{file=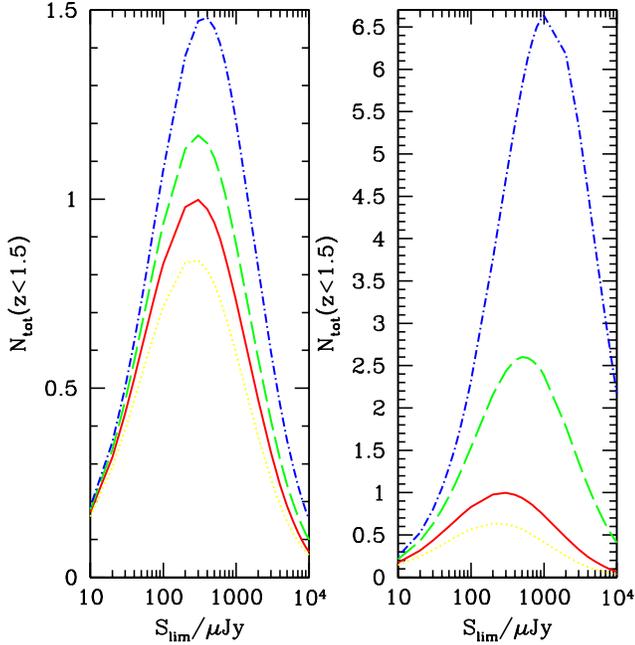,width=9.0cm,height=9.0cm}}
%\vskip -1cm
\caption{The effect of varying the cluster model parameters $\alpha$ and $\beta$ for $\theta_{\rm FWHM}=1^{\prime}$ and $f=30{\rm GHz}$. On the left we vary $\alpha$, while keeping $\beta=2/3.$ The dotted line is $\alpha=5$, solid line is $\alpha=10$, dashed line is $\alpha=50$ and dot-dashed line is $\alpha=100$. On the right we vary $\beta$, with $\alpha=10$. The dotted line is $\beta=1/3$, the solid line is $\beta=2/3$, dashed line is $\beta=1$ and dot-dashed line is $\beta=4/3$. In each case the curves are normalized in such a way that the maximum value of $N_{\rm tot}$ for $\alpha=10$ and $\beta=2/3$ is one. The solid lines are the equivalent to that in the bottom right of Fig.~\ref{fig:ntot2}. We see that the optimum value of $S_{\rm lim}$ is relatively independent of $\alpha$, but appears to change significantly as we vary $\beta$. In particular the optimum value is $S_{\rm lim}=280\mu{\rm Jy}$ for $\beta=2/3$ and is $\sim 1000\mu{\rm Jy}$ for $\beta=4/3$.}
\label{fig:ntot_alpha}
\end{figure}

We have also varied the parameters of the isothermal $\beta$-model, $\alpha$ and $\beta$. For a general value of $\beta$ the profile function is given by 
\be 
\zeta(\theta)=\left(1+{\theta^2\over \theta_{\rm c}^2}\right)^{{1\over 2}-{3\over 2}\beta}{J\left[\left({\alpha^2-\theta^2/\theta_{\rm c}^2\over 1+\theta^2/\theta_{\rm c}^2}\right),\beta\right]\over J(\alpha,\beta)}\,,
\ee
where 
\be
J(a,b)=\int^{a}_{0}(1+x^2)^{-{3b\over 2}}\,dx\,.
\ee
Fig.~\ref{fig:ntot_alpha} shows the effect of varying $\alpha=5$, 10, 50 and 100 and $\beta=1/3$, 2/3, 1 and 4/3 for a survey with $\theta_{\rm FWHM}=1^{\prime}$ and $f=30{\rm GHz}$. We see that the optimal value of $S_{\rm lim}$ varies only by about a factor of 2 when $\alpha$ varies by a factor of 20  between 5 and 100. We note also that larger values of $\alpha$ lead to more clusters detected; about a factor of 2 between $\alpha=5$ and 100. This is because for larger values of $\alpha$ the clusters are more point-like, making a given mass easier to detect. 

The variations as a function of $\beta$ are more significant. The optimal value of $S_{\rm lim}$ for $\beta=2/3$ is $280\mu{\rm Jy}$ (see Table 1) whereas for $\beta=4/3$ it is $\approx 1{\rm mJy}$. Moreover the total number of clusters increases by a factor $\sim 10$ in going from $\beta=1/3$ to $\beta=4/3$. As before this is because as $\beta$ increases the cluster becomes more point-like: for $\beta=1/3$, the electron number density $n_{\rm e}\propto r^{-1}$, whereas for $\beta=4/3$, $n_{\rm e}\propto r^{-4}$.

\subsection{Mass function}

The final major aspect of our calculation which could contain systematic errors is the comoving number of density of objects $dn/dM$ which is computed from N-body simulations~\cite{EVRARD}. It is clear that the simulations used, while the best available at present, only have limited dynamical range, and the fitted formula for $dn/dM$ could be inaccurate at the extreme values of $M$, both large and small. In order to investigate the possible effects of this on our calculation we have tried three alternative mass functions in addition to the one used in the rest of the paper. 

\begin{figure} 
\setlength{\unitlength}{1cm}
%\vskip -2cm
\centerline{\psfig{file=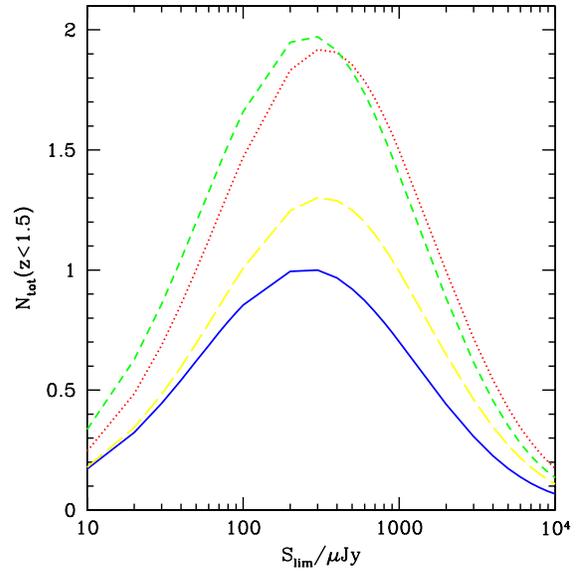,width=8.0cm,height=8.0cm}}
%\vskip -1cm
\caption{The effect of different mass functions for $\theta_{\rm FWHM}=1^{\prime}$ and $f=30{\rm GHz}$. The solid line is the standard case of the Evrard et al (2002) mass function used in the earlier discussion, notably Fig.~\ref{fig:ntot2}. All the other surveys have been normalized to the number of objects which would be found relative to this mass function. The dotted line is for Jenkins et al (2001), long dashed line is for ST with $M=M_{60}$ and the short-dashed line is for PS with $M=M_{200}$. The optimum value appears to be relatively independent of the mass function, but the overall number of clusters varies by about a factor of 2.}
\label{fig:ntot_mass}
\end{figure}

In Fig.~\ref{fig:ntot_mass} we have plotted the same quantity as in Fig.~\ref{fig:ntot2} for $\theta_{\rm FWHM}=1^{\prime}$ and $f=30{\rm GHz}$ using the  mass function already used~\cite{EVRARD} which is based on VIRGO consortium simulations, their alternative mass function~\cite{jenk}, the ST~\cite{ST} mass function using $M=M_{60}$ and the PS~\cite{PS} mass function using $M=M_{200}$. We see that the maximum of the curve, and hence the optimal survey depth is almost independent of the mass function within the bounds of those functions which we have investigated. Rather worryingly, the overall number of clusters found for a given flux limit varies by a factor of two for the different mass functions. This point was highlighted in Battye \& Weller (2003), and it is clearly something which needs further investigation. Nonetheless, we conclude that uncertainties in the mass function are not significant when considering the optimal depth of a survey.

\section{Conclusions}
\label{concl}

We have discussed in detail the optimal depth of an SZ cluster survey and have as a by-product gained some insight in the optimal experimental configuration. Ignoring the effects of confusion noise, for every beamsize considered we have shown that there  exists an optimum depth at which one will find the largest  number of clusters for a fixed integration time. The number of clusters found as a function of the flux density or temperature limit is typically quite broad, which is fortunate since the calculations we have made contain many uncertainties. We have shown that the most significant of these are those relating to the $M-T$ relation and the cluster profile. Even within the known uncertainties in these parameters it should be possible to get within $50\%$ of the optimum.

For an instrumental setup with $\theta_{\rm FWHM}^{\rm min}<2^{\prime}$ and assuming that beam smoothing is possible, one is brightness temperature limited and it is inefficient to make very deep surveys searching for clusters which match a much smaller beam. One is always in a position to perform shorter integrations on many small beam areas and smooth them down to the lower resolution where the maximum number of clusters are available. These conclusions are particularly sensitive to some of the more sophisticated experimental techniques which are used to counteract the effects of the atmosphere. We have discussed two: in the case of beam switching, we have shown that one is not severely affected if the beam switch is not so small as to prevent one detecting the most abundant clusters which are which have a virial diameter of $\sim 2^{\prime}$.

 The discussion of the optimal design of an interferometer and the survey strategy is complicated by the extra freedom due to density of visibilities in the $u$-$v$ plane. We have discussed how our methods can be applied in that case, but, other than asserting that the most filled interferometer would have the optimal design, we have not made any statements on the optimal strategy. Given a particular situation it might be possible to do this in more detail.

We hope that our calculations can be used to assist observers in their choice of survey strategy. We acknowledge numerous times that our predictions are model dependent. One could imagine a method by which one could search adaptively for the optimum depth of a survey as part of the process of performing the observation. The basic idea is to take some small patch of sky, compared to the potential total survey volume, and compute the number of sources as a function of $S_{\rm lim}$. When the logarithmic gradient of this quantity is greater than -2, the optimum has been achieved, and that value of $S_{\rm lim}$ should be used on the rest of the survey area. Clearly, this will only be possible in scenarios where one is expected to find a large number of objects since one will have to find a sufficiently large number in the initial test area to make the estimate of the gradient accurate enough. 

\subsection*{ACKNOWLEDGEMENTS}

RAB is funded by PPARC and thanks the members of the CMB group at
Jodrell Bank Observatory, in particular Peter Wilkinson, Ian Browne
and Clive Dickinson, for many useful discussions on the content of the
is paper. JW is supported by the DOE and the NASA grant NAG 5-10842 at
Fermilab. 

\subsection*{Appendix A}

Let us consider computing the integral ${\cal I}(B,\zeta)$ when $\zeta(\theta)=F(y)$ where $y=\theta^2/\theta_{\rm vir}^2$ for $\theta\le\theta_{\rm vir}$ and zero for $\theta>\theta_{\rm vir}$. The function is defined so that  $F(0)=1$. It is easy to see that for the isothermal $\beta$-model with $\beta=2/3$ 
\be 
F(y)=\left(1+\alpha^2y^2\right)^{-1/2}{\tan^{-1}\left[{\alpha(1-y^2)^{1/2}\over 1+\alpha^2y^2}\right]\over \tan^{-1}\alpha}\,,
\ee
and it is likely that most sensible profile functions will have this property.

If we define $R=4(\theta_{\rm vir}/\theta_{\rm FWHM})^2\log 2$ then 
\be 
{\cal I}(B,\zeta)={\int_0^1\,e^{-Ry}\,F(y)\,dy\over \int_0^1\,F(y)\,dy}\,.
\ee
One can see that in the limit $R\rightarrow 0$ the function ${\cal I}\rightarrow 1$ which corresponds to the point source limit. The opposite limit, $R\rightarrow\infty$, is more interesting since it corresponds to that when the cluster is much larger than the beam. One can show that the leading order term in the asymptotic expansion in $1/R$ is 
\be
{\cal I}\sim \left[R\int_0^1\,F(y)\,dy\right]^{-1}\propto \theta_{\rm vir}^{-2}\,.
\ee
From (\ref{smt}) we see that $S\propto \langle T_{\rm e}\rangle_{\rm n}M_{\rm vir}{\cal I}d_{\rm A}^{-2}$ and since $R_{\rm vir}=\theta_{\rm vir}d_{\rm A}$ we can deduce that 
\be 
S\propto {\langle T_{\rm e}\rangle_{n}M_{\rm vir}\over R_{\rm vir}^2}\,,
\ee
which is finite as $z\rightarrow 0$. Hence, we see that the limiting mass tends to a finite value at $z=0$ in contrast to the case of a point source.

\subsection*{Appendix B}

Let us consider ${\cal I}(B_{\rm eff},\zeta)$ for $B_{\rm eff}(\theta)$ given by that for a switch beam system (\ref{switch}). We will assume that $F(y)$ and $R$ are defined  as in Appendix A, and will define $Q=4(\theta_{\rm s}/\theta_{\rm FWHM})^2\log 2$. Using these definitions we find that 
\be
{\cal I}(B_{\rm eff},\zeta)={\int_0^1\,e^{-Ry}\left[1-e^{-Q}I_0\left(2\sqrt{RQy}\right)\right]\,F(y)\,dy\over \int_0^1\,F(y)\,dy}\,.
\ee
Understanding the asymptotic behaviour of this integral is more difficult than in the standard case. In order to make some progress let us write $F(y)$ as a power series 
\be 
F(y)=\sum_{n=0}^{\infty}A_{m}y^{m}\,.
\ee
where $A_0=1$ and $A_1<0$. From this we can deduce that  
\be 
{\cal I}(B_{\rm eff},\zeta)={\sum_{m=0}^{\infty}A_{m}\left(B_{m}-e^{-Q}C_{m}\right)\over \int_0^1F(y)dy}\label{ll}
\ee
where 
\begin{eqnarray}
B_m&=&\int_0^1e^{-Ry}y^mdy\,, \\
C_m&=&\int_0^1e^{-Ry}I_0\left(2\sqrt{RQy}\right)y^mdy\,.
\end{eqnarray}
One can write $C_{m}$ in terms of $B_{m}$ by writing $I_0(x)$ as a power series in $x$,
\be
C_{m}=\int_{n=0}^{\infty}{1\over (n!)^2}(RQ)^{n}B_{m+n}\,.
\ee
Moreover, it can be shown that $B_m\sim m!/R^{m+1}$ and hence $C_0\sim e^{Q}/R$ and $C_1\sim (Q+1)e^{Q}/R^2$. From this we see that in the asymptotic expansion, the first term in (\ref{ll}) cancels and therefore
\be
{\cal I}(B_{\rm eff},\zeta)\sim {-A_1Q\over R^2\int_0^1F(y)dy}\,.
\ee
This cancellation is enough to ensure that ${\cal I}\rightarrow0$ as $z\rightarrow 0$  sufficiently fast so that 
\be
S\propto {\langle T_{\rm e}\rangle_nM_{\rm vir}d_{\rm A}^2\over R_{\rm vir}^4}\,,
\ee
in this limit. It is easy to see that this forces the limiting mass to tend to infinity as $z\rightarrow 0$.

\bibliographystyle{mnras}

\label{lastpage}
\end{document}